%% file: main.tex
\documentclass[letterpaper,twocolumn,10pt]{article}
\usepackage{usenix-2020-09}
\usepackage{authblk}
\usepackage{amsmath}
\usepackage[final]{changes}

\usepackage{graphicx} 
\usepackage{float} %设置图片浮动位置的宏包
\usepackage{subfigure} %插入多图时用子图显示的宏包
\usepackage{hyperref}
\makeatletter
\def\UrlAlphabet{%
      \do\a\do\b\do\c\do\d\do\e\do\f\do\g\do\h\do\i\do\j%
      \do\k\do\l\do\m\do\n\do\o\do\p\do\q\do\r\do\s\do\t%
      \do\u\do\v\do\w\do\x\do\y\do\z\do\A\do\B\do\C\do\D%
      \do\E\do\F\do\G\do\H\do\I\do\J\do\K\do\L\do\M\do\N%
      \do\O\do\P\do\Q\do\R\do\S\do\T\do\U\do\V\do\W\do\X%
      \do\Y\do\Z}
\def\UrlDigits{\do\1\do\2\do\3\do\4\do\5\do\6\do\7\do\8\do\9\do\0}
\g@addto@macro{\UrlBreaks}{\UrlOrds}
\g@addto@macro{\UrlBreaks}{\UrlAlphabet}
\g@addto@macro{\UrlBreaks}{\UrlDigits}
\definechangesauthor[name={skw}, color=orange]{skw}
\definechangesauthor[name={wch}, color=blue]{wch}
\definechangesauthor[name={gml}, color=magenta]{gml}
\definechangesauthor[name={ckd}, color=red]{ckd}
\definechangesauthor[name={zmm}, color=purple]{zmm}

\makeatother
\usepackage{threeparttable}
\usepackage{multirow} %合并多行单元格的宏包
\usepackage{booktabs} %s三线表格中的上中下直线线型设置宏包，在这个包中水平线被教程\toprule、midrule、buttomrule。
\newcommand{\tabincell}[2]{\begin{tabular}{@{}#1@{}}#2\end{tabular}}
\usepackage{amssymb}
\usepackage{color}
\usepackage{colortbl}
\usepackage{authblk}

\begin{document}
\pagestyle{empty}  % no page number for the second and the later pages
\thispagestyle{empty} % no page number for the first page
\title{\Large \bf Weak Links in Authentication Chains: \\ A Large-scale Analysis of Email Sender Spoofing Attacks
}

\author{
Kaiwen Shen $^{1,*}$,
Chuhan Wang $^{1,}$ \thanks{ Both authors contributed equally to this work.},
Minglei Guo $^1$, 
Xiaofeng Zheng $^{1,2,\dag}$,
Chaoyi Lu $^1$, 

Baojun Liu $^{1,}$\thanks{
 Corresponding authors:\{zxf19, lbj15\}@mails.tsinghua.edu.cn.},
Yuxuan Zhao $^4$, 
Shuang Hao $^3$, 
Haixin Duan $^{1,2}$,
Qingfeng Pan $^5$
and Min Yang $^6$
}

\affil{
    $^1$Tsinghua University
    $^2$Qi An Xin Technology Research Institute
    $^3$University of Texas at Dallas
    
    $^4$North China Institute of Computing Technology
    $^5$Coremail Technology Co. Ltd
    $^6$Fudan University
}

\renewcommand\Authands{ and }

\maketitle

\begin{abstract}
\input{0_abstract.tex}
\end{abstract}

\input{1_intro.tex}

\input{2_background.tex}

\input{3_attack_model_and_experiments.tex}

\input{4_attacks.tex}

\input{5_combined_attacks.tex}

\input{6_discussion_mitigation.tex}

\input{7_disclosure_and_response.tex}

\input{8_related_work}
\input{9_conclusion}
\input{10_acknowledgments.tex}

\bibliographystyle{plain}
\bibliography{paper}

% \appendix
% \input{appendix.tex}

\end{document}

%% file: 0_abstract.tex
As a fundamental communicative service, email is playing an important role in both individual and corporate communications, which also makes it one of the most frequently attack vectors.
An email's authenticity is based on an authentication chain involving multiple protocols, roles and services, the inconsistency among which creates security threats.  
\replaced[id=skw]{Thus, it depends on the weakest link of the chain, as any failed part can break the whole chain-based defense.}{
The chain-based authentication structure means a failure in any part can break the whole chain. }

This paper systematically analyzes the transmission of an email and identifies a series of new attacks capable of bypassing SPF, DKIM, DMARC and user-interface protections.
In particular, by conducting a "cocktail" joint attack, more realistic emails can be forged to penetrate the celebrated email services, such as Gmail and Outlook. 
We conduct a large-scale experiment on 30 popular email services and 23 email clients, and find that all of them are vulnerable to certain types of new attacks.
We have duly reported the identified vulnerabilities to the related email service providers, and received positive responses from 11 of them, including Gmail, Yahoo, iCloud and Alibaba.
Furthermore, we propose key mitigating measures to defend against the new attacks.
Therefore, this work is of great value for identifying email spoofing attacks and improving the email ecosystem's overall security.

%% file: 1_intro.tex
\section{Introduction}
\label{sec:intro}

% As the value of data grows, the abundant individual and corporate information contained in the email has made email a key target of cyber attacks~\cite{jagatic2007social}.
Email service has been a popular and essential communicative service with abundant individual and corporate information, which makes it a key target of cyber attacks~\cite{jagatic2007social}.
Yet, the email transmission protocols are far from capable of countering potential attacks.
% The security of an email system relies on a multi-party trust chain mechanism maintained by various email services, the complexity of which has increased the systemic vulnerability to cyber attacks.
An email system's security relies on a multi-party trust chain maintained by various email services, which increases its systemic vulnerability to cyber attacks.
% Specifically, the problem of this chain-based authentication structure is that any failed part can break the whole authentication chain. The email authentication chain involves multiple protocols, roles, and services.

As the Wooden Bucket Theory reveals, a bucket's capacity is determined by its shortest stave. 
The authenticity of an email depends on the weakest link in the authentication chain.
Even a harmless issue may cause unprecedented damages when it is integrated into a more extensive system. 
Generally, the email authentication chain involves multiple protocols, roles and services, any failure among which can break the whole chain-based defense.

First, despite the existence of various security extension protocols (e.g., SPF\cite{kitterman2014sender}, DKIM\cite{allman2007domainkeys} and  DMARC\cite{kucherawy2015domain}) to identify spoofing emails, 
\replaced[id=skw]{spoofing attacks might still succeed due to the inconsistency of entities protected by different protocols. }{the inconsistency of entities protected by different protocols still makes it possible for spoofing attacks to succeed.}

Second, authentication of an email involves four different roles: senders, receivers, forwarders and UI renderers. 
\replaced[id=skw]{Each role should take different security responsibilities. }{Different roles should bear different security responsibilities.}
If any role fails to provide a proper security defensive solution, an email’s authenticity can not be guaranteed.

Finally, security mechanisms are implemented by different email services with inconsistent processing strategies.
% Their processing strategies could be inconsistent with each other.
Besides, those security mechanisms  are implemented by different developers, some of which deviate from RFC specifications while dealing with emails with ambiguous headers. 
Therefore, there are a number of inconsistencies among different services. 
Attackers can utilize these inconsistencies to bypass the security mechanisms and present deceptive results to the webmails and email clients.

% Overall, as the Wooden Bucket Theory reveals, a bucket's capacity is determined by its shortest stave.
% The authenticity of email depends on the weakest link in the email authentication chain. 
% A harmless issue may cause unprecedented damages when it is integrated into a more extensive system. 
% Similarly, a "cocktail" spoofing attack that combines several different attacks could pose a huge threat to email ecosystem.
% Similarly, a "cocktail" joint attack could pose a huge threat to email ecosystem.

This work systematically analyzes four critical stages of authentication in the email delivery process: sending authentication, receiving verification,  forwarding verification and UI rendering. 
%Furthermore, 
We found 14 email spoofing attacks capable of bypassing SPF, DKIM, DMARC and user-interface protections. 
By combining different attacks, a spoofing email can completely pass all prevalent email security protocols, and no security warning is shown on the receiver's MUA. 
%Therefore, 
We show that it is still challenging to identify whether such an email is spoofing, even for people with a senior technical background.

To understand the real impacts of spoofing email attacks in the email ecosystem, we conducted a large-scale experiment on 30 popular email services with billions of users in total. 
Besides, we also tested 23 popular email clients on different operating systems to measure the impact of attacks on the UI level. 
All of them are vulnerable to certain types of attacks, including reputable email services, such as Gmail and Outlook. 
\replaced[id=skw]{
%Moverover, 
We have already duly reported all identified issues to the involved email service providers and received positive responses from 11 of them (e.g., Gmail, Yahoo, iCloud, Alibaba Cloud). }{
All identified issues have already been duly reported to the involved email service vendors. 
We have received positive responses from 11 email vendors (e.g., Gmail, Yahoo, iCloud, Alibaba Cloud) so far. }

Our work shows the vulnerability of the chain-based authentication structure in the email ecosystem.
\replaced[id=skw]{The attacks reveal that more security issues are led by the inconsistency among multiple parties' understanding and implementation of security mechanisms.
}{
The attacks we found share the high-level theme that the inconsistency among multiple parties' understanding and implementation of security mechanisms has led to more security issues. }
\replaced[id=skw]{To counter email spoofing attacks, we proposed a UI notification scheme.}{Based on our findings, this work have proposed a UI notification scheme to counter email spoofing attacks. }
Coremail, a well-known email service provider in China, has adopted our scheme and implemented it on the webmails and email clients for users. 
Besides, we have also released our testing tool on Github for email administrators to evaluate and \replaced[id=skw]{increase}{strengthen} their security.

% \myparagraph{Contributions.}
\noindent \textbf{Contributions.} 
To sum up, we make the following contributions:
\begin{itemize}
	\item By analyzing the email authentication chain systematically, we identified a total of 14 email spoofing attacks, 9 of which (i.e., A$_3$, A$_6$, A$_7$, A$_8$, A$_9$, A$_{10}$, A$_{11}$, A$_{13}$, A$_{14}$) are new attacks, to the best of our knowledge so far. By combining different attacks, we can forge more realistic spoofing email to penetrate celebrated email services like Gmail and Outlook.
	\item We conducted a large-scale measurement on 30 popular email services and 23 email clients. We found all of them are vulnerable to some of attacks. We have responsibly disclosed  vulnerabilities and received positive responses from 11 email vendors (e.g., Gmail, Yahoo, iCloud and Alibaba Cloud).
	\item To enhance the protection of email system against spoofing attacks, we\deleted[id=skw]{ have proposed some key mitigation measures to defend against email spoofing attacks. We} proposed a UI notification scheme and provided an email security evaluation tool for email administrators to evaluate and \replaced[id=skw]{increase}{strengthen} their security.
\end{itemize}

%% file: 2_background.tex
\section{Background}
\label{sec:background}

\subsection{Email Delivery Process}
\begin{figure}[h]
    \centering
    \includegraphics[width=7cm]{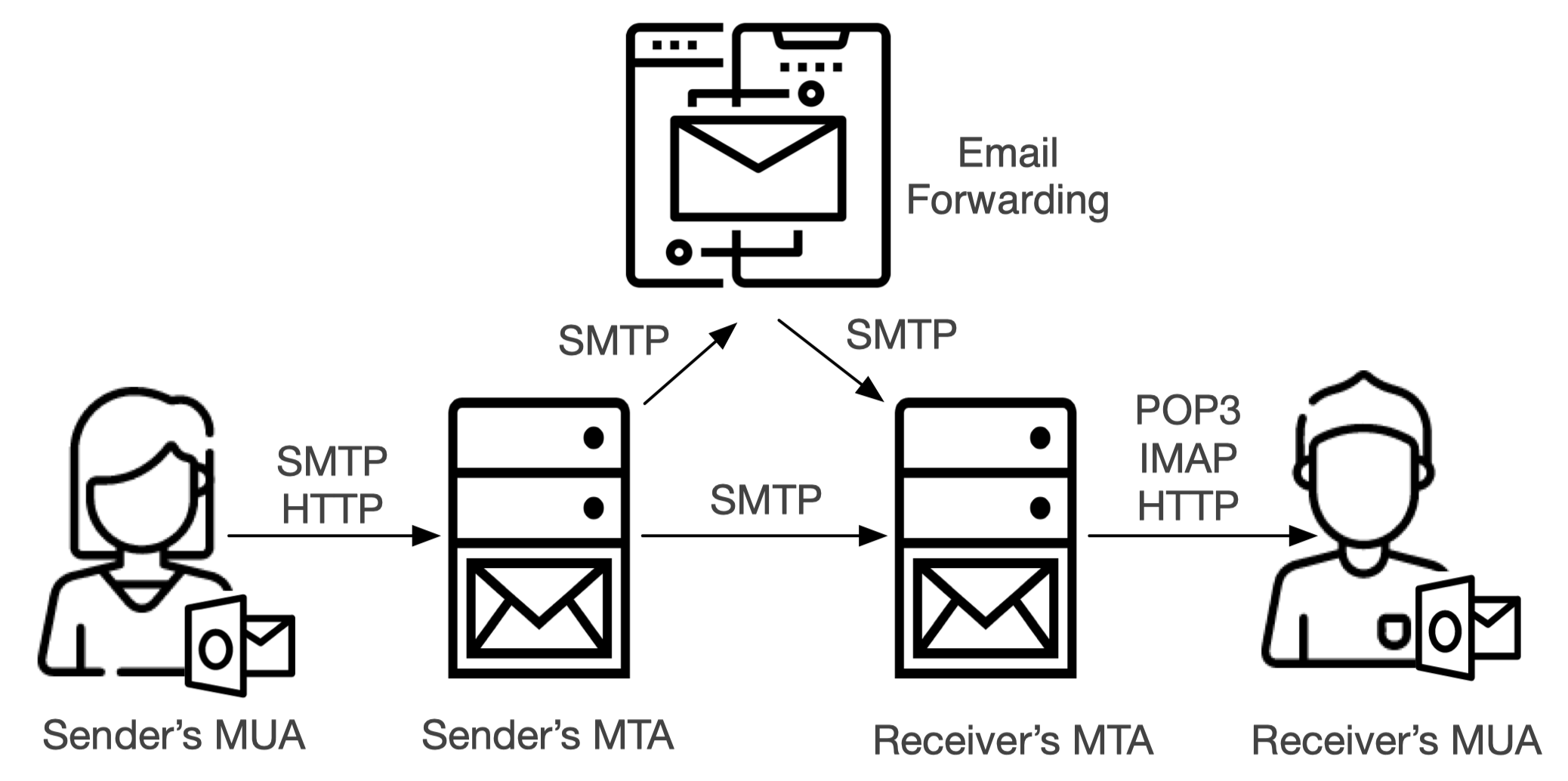}
    \vspace{0.02cm}
    \caption{The email delivery process.}
    \label{fig:email_transmission}
    \end{figure}

Simple Mail Transfer Protocol (SMTP)\cite{postel1982simple} is a basic protocol for email services. 
Figure~\ref{fig:email_transmission} shows the basic email delivery process.
An email written by a sender is transmitted from the Mail User Agent (MUA) to the Mail Transport Agent (MTA) via SMTP or HTTP protocol. 
Then, the sender's MTA transmits the email to the receiver's MTA via the SMTP protocol, which later delivers the email content to the receiver's MUA via HTTP, IMAP or POP3\cite{klensin1997imap} protocols.

Extra transmission needs could complicate the actual delivery process. When the original email's target recipient is a mailing list or configured with an automatic email forwarding service, the email will be relayed through an email server, such as the email forwarding server in Figure~\ref{fig:email_transmission}. The email forwarding server will modify the receiver's address and redeliver it.
    
In the SMTP communication process, a sender's identity information is contained in multiple fields in a complex manner. \label{multiple_identity}(1) \texttt{Auth username}, the username used in the \texttt{AUTH} command to authenticate the client to the server. (2) \texttt{MAIL From}, the sender on the envelope, is mainly used for identity verification during the email delivery process. (3) \texttt{From}, the sender in the email body, is the displayed address that the email client shows to the user.
%, so this field is also designed to be more flexible. 
(4) \texttt{Sender}, the \texttt{Sender} field is used to identify the real sender when there are multiple addresses in the \texttt{From}. 
The inconsistency of these fields provides the basis for email spoofing attacks.

As shown in Figure~\ref{fig:email_transmission}, the authentication in the email transmission process involves four important stages. 

\noindent \textbf{Email Sending Authentication. } When sending an email from the MUA via the SMTP protocol, the sender needs to enter his username and password for authentication. In this part, the sender's MTA not only needs to verify the user's identity but also to ensure the \texttt{Mail From} is consistent with the \texttt{Auth username}.

\noindent \textbf{Email Receiving Verification. } When the receiver’s MTA receives the email, MTA validates the sender’s authenticity through SPF, DKIM and DMARC protocols. See Section \ref{Email Security Extension Protocols} for details of these protocols. 

\noindent \textbf{Email Forwarding Verification. } Email automatic forwarding is another commonly used way to send emails. When a forwarder automatically forwards an email, it should verify the sender's address. If the DKIM signature is enabled, the original DKIM verification status should be "pass" at first, then a new DKIM signature will be added. If the ARC\cite{blank2019authenticated} protocol is deployed, the ARC verification chain will also be verified. 

\noindent \textbf{Email UI Rendering.} This stage is to provide users with a friendly email rendering display. Unfortunately, most popular email clients' UI will not present the authenticity check result to users. Some encoding formats or special characters can mislead receiver with a spoofing address. We argue that Email UI rendering is the last but crucial step in the authentication process, which is often overlooked in previous research.

\subsection{Email Spoofing Protections}

    \begin{figure}[t]
    \centering
    \includegraphics[width=7cm]{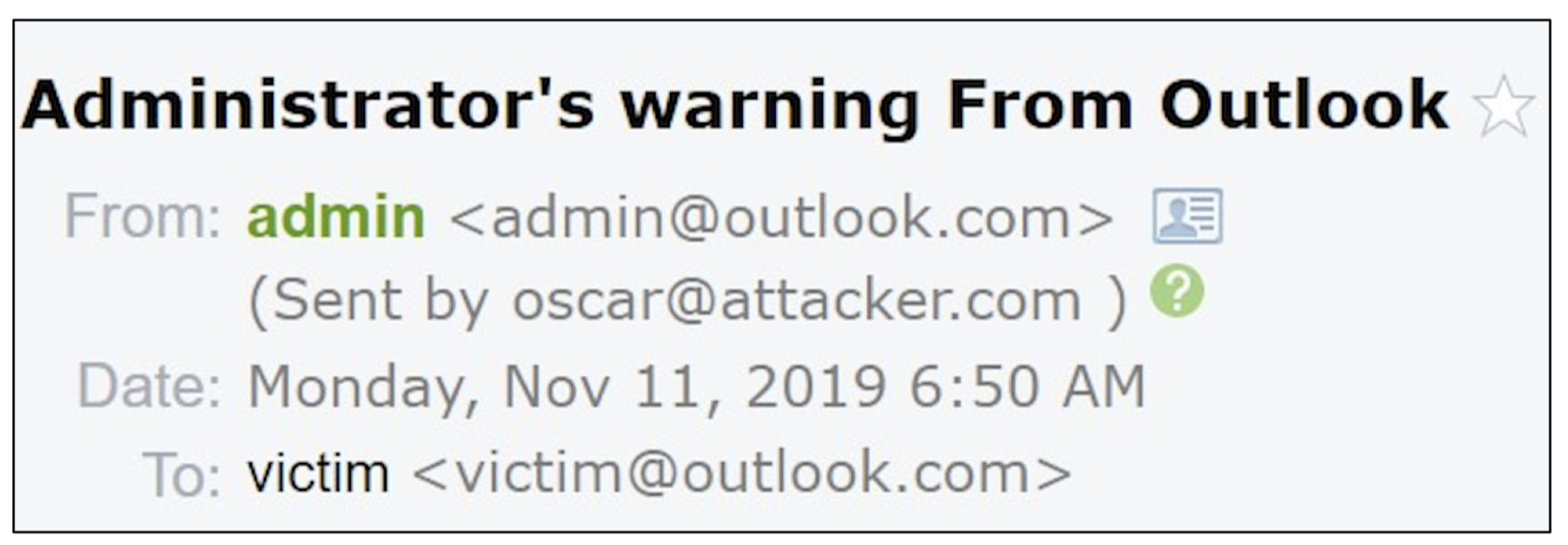}
    \caption{A spoofing email that fails the Sender Inconsistency Checks. 
    % "Sent by" indicates the email 
}
    \label{fig:sender-check}
    \end{figure}
\subsubsection{Email Security Extension Protocols\label{Email Security Extension Protocols}}
To defend against email spoofing attacks, various security extensions have been proposed and standardized. At present, SPF, DKIM and DMARC protocols are the most widely used ones.

\noindent \textbf{SPF.}	Sender Policy Framework (SPF)\cite{kitterman2014sender} is an IP-based authentication protocol. It marks and records the sender's domain and IP address together. The receiver can determine whether the email is from the claimed domain by querying the SPF record under the DNS server corresponding to the sender's domain name.

\noindent \textbf{DKIM.} DomainKeys Identified Mail (DKIM)\cite{crocker2011domainkeys} is an authentication protocol based on digital signatures. It uses an asymmetric key encryption algorithm to allow a sender to add a digital signature to an email's header to identify spoofing attempts during transmission. The receiver can retrieve the sender's public key from DNS querying to verify the signature, and then determine whether the email was spoofing or modified.

\noindent \textbf{DMARC.} Domain-based Message Authentication, Reporting and Conformance (DMARC)\cite{kucherawy2015domain} is an authentication system based on the results of SPF and DKIM verification. It introduces a mechanism for multiple authenticated identifiers alignment, which associates the identity information in \texttt{From} with the authenticated identifier of SPF or DKIM. Meanwhile, the domain owner can publish a policy suggesting solutions to the recipient to handle unverified emails sent by this domain name. The domain owner can get regular feedback from the recipient. Specifically, DMARC employs an "or" status check of the SPF and DKIM verification results. If an email passes the detection of either SPF or DKIM, and \texttt{From} can be aligned with the authenticated identifier, it passes the validation of DMARC.

\subsubsection{UI-level Spoofing Protections}

UI rendering is a crucial part that affects the users’ perception of an email's authenticity. \deleted{Many vendors have been aware of this fact and are cautious about their UI level design.} However, the necessity of increasing UI level protection has not yet fostered any prevalent security protocol. Each Email vendor employs different UI level protections, and there is no widely accepted comprehensive protection mechanism so far.

\noindent \textbf{Sender Inconsistency Checks (SIC\label{SIC}).}
As shown in Figure \ref{fig:sender-check}, some email services add a security indicator to alert the receiver that the actual sender (\texttt{MAIL From}) may not be the displayed one (\texttt{From}). 
It is worth noting that this inconsistency exists throughout the email system, including email forwarding, alias, and email subscriptions. Therefore, the receiver's MTA cannot directly reject an email because of the inconsistency, which lowers the success rate to detect spoofing emails.
However, the protection measure addressing this issue has not received a clear definition in the industry yet. We define this protection measure as the Sender Inconsistency Checks (SIC).

%% file: 3_attack_model_and_experiments.tex
 \section{Attack Model and Experiments}
 \label{sec:attack-model}

\subsection{Attack Model}

\begin{figure}[t]
\centering
\includegraphics[width=7cm ]{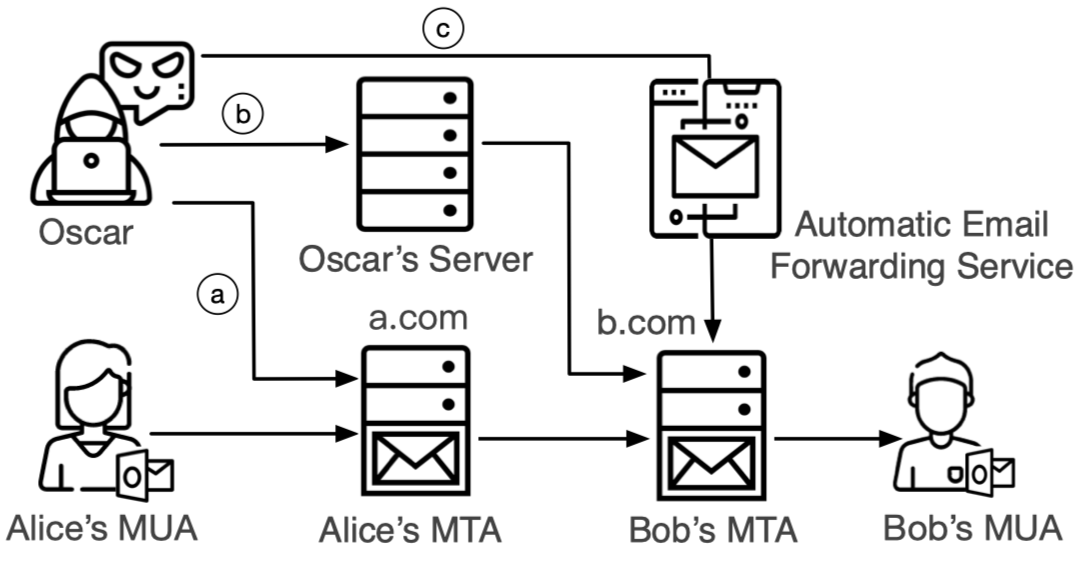}
\caption{The Attack Model: \textcircled{a}, \textcircled{b} and \textcircled{c} represent shared MTA Attack, Direct MTA Attack and Forward MTA Attack respectively.}
\label{fig:attack-model}
\end{figure}

As shown in Figure ~\ref{fig:attack-model}, the attack model of email spoofing attacks includes a trusted email sender (Alice, which has an email account under a.com), a victim receiver (Bob, which has an email account under b.com), and an adversary (Oscar).
Specifically, Oscar’s goal is to send an email to Bob, spoofing Alice@a.com and bypassing all security validation. 
 
In general, there are three common types of email spoofing attacks.

\noindent \textcircled{a} 
\textbf{Shared MTA Attack.}
\replaced[id=skw]{We assume that Oscar has an email account (Oscar@a.com), which is different from Alice's account (Alice@a.com). Oscar can send spoofing emails through the MTA of a.com by modifying the \texttt{Mail From}/ \texttt{From}/ \texttt{Auth username} headers.}{Oscar can send spoofing emails through the MTA of a.com, by which he can send spoofing emails through the MTA of a.com. }
\replaced[id=skw]{Since}{Because} the credibility of the sender's MTA IP is an essential factor affecting the spam engine's decision algorithm\cite{blanzieri2008survey}, 
\replaced[id=skw]{the spoofing email can easily enter the victim's inbox.}{it is easy for Oscar's spoofing email to enter the victim's inbox. }
The \added[id=skw]{IP of the} sender's MTA is in a.com's SPF scope. The sender's MTA \replaced[id=skw]{may}{will} also automatically \replaced[id=skw]{attach}{add} DKIM signatures to the spoofing email\deleted[id=skw]{ if SPF and DKIM are enabled}. 
Therefore, Oscar has little difficulty in bypassing the SPF/DKIM/DMARC verification and spoofs Alice@a.com.

\noindent \textcircled{b}
\textbf{Direct MTA Attack.}
\added[id=skw]{Oscar can also send spoofing emails through his own email server.}
\replaced[id=skw]{Note that the communication process between the sender's MTA and the receiver's MTA does not have an authentication mechanism. Oscar can }{Because the communication process between the sender's MTA and the receiver's MTA does not have an authentication mechanism, attackers can }spoof an arbitrary sender by specifying the \texttt{Mail From} and the \texttt{From} headers.
This attack model can ensure all spoofing emails reach the receiver's MTA without being influenced by the strict sending check \added[id=skw]{of the sender's MTA}.

\noindent \textcircled{c}
\textbf{Forward MTA Attack.}
Oscar can abuse the email forwarding service to send spoofing emails. First of all, Oscar can send a spoofing email to Oscar@a.com\added[id=skw]{, an email account belonging to Oscar on the forwarding email service.}
Next, he can configure the forwarding service to automatically forward this spoofing email to the victim (Bob@b.com).
This attack model has three major advantages. 
First, this attack has the same advantages as the Shared MTA attack mode because the receiver's MTA (b.com) believes that the emails come from the legitimate MTA (a.com).
Moreover, this attack can also bypass the strict sending check \added[id=wch]{of the sender's MTA} (e.g., a mismatch between \texttt{Mail From} and \texttt{From} headers). \added[id=skw]{Finally, the forwarding service may give the forwarded email a higher security endorsement (e.g., adding a DKIM signature that shouldn't be added).}

As such, the sender authentication issues can occur in four stages, including sending authentication, receiving verification, forwarding verification and UI rendering, which can all pose potential security threats.

Further, we define the goals of a successful attack as follows: 
(1) the receiver's MUA incorrectly renders the sender address as it comes from a legitimate domain name, rather than the attacker's real one; 
(2) the receiver's MTA incorrectly verifies the sender of spoofing emails; 
(3) the receiver's MUA does not display any security alerts for spoofing emails.

\begin{figure}[t]
\centering                                                          
\subfigure[Gmail's Web UI does not display any spoofing alerts]{      
\begin{minipage}{8cm}
\centering                                               
\includegraphics[width=7cm]{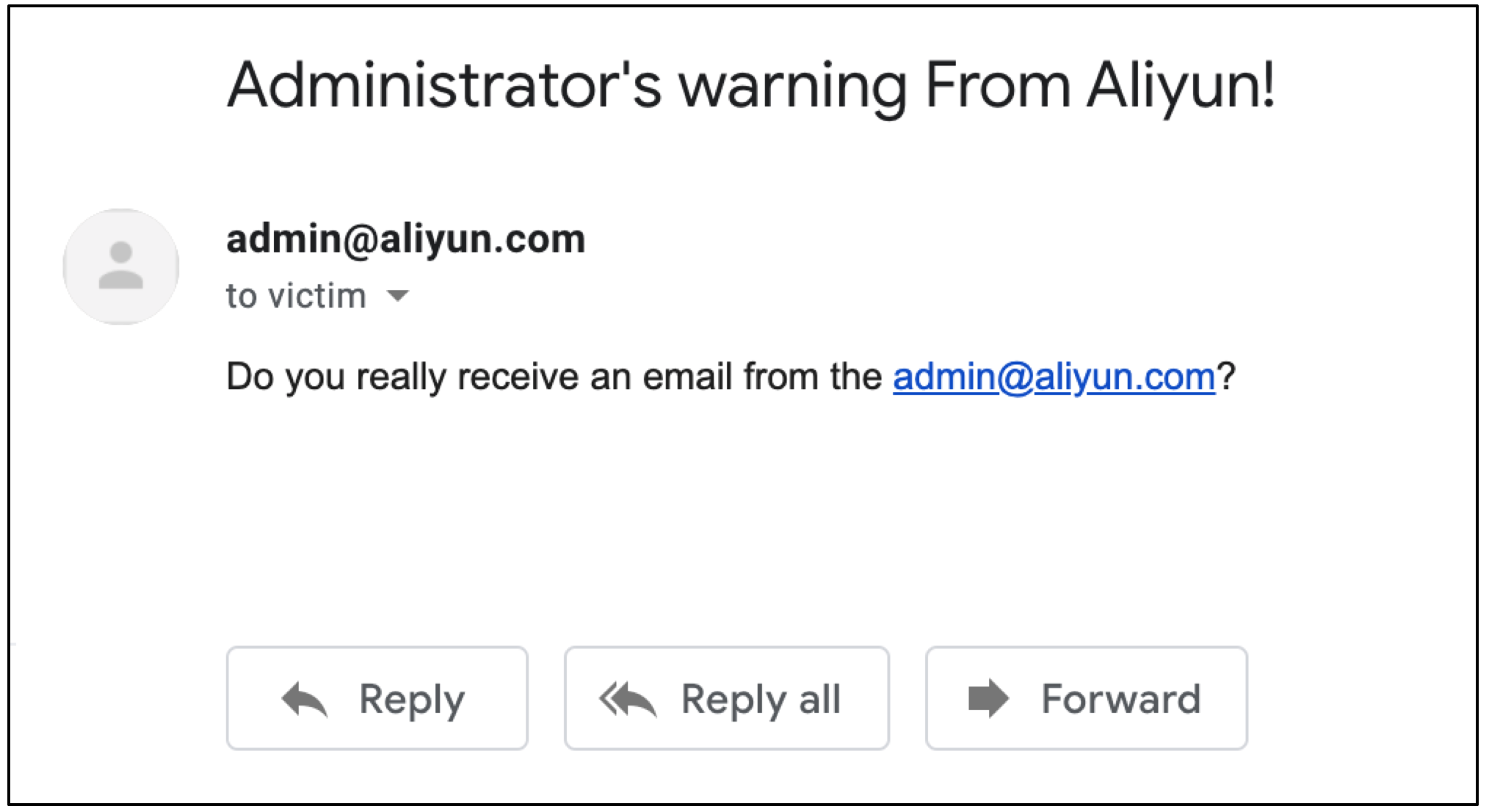}               
\end{minipage}
} 
\quad
\subfigure[The spoofing email passes all email security protocol verification]{
\begin{minipage}{8cm}
\centering         
\includegraphics[width=7cm]{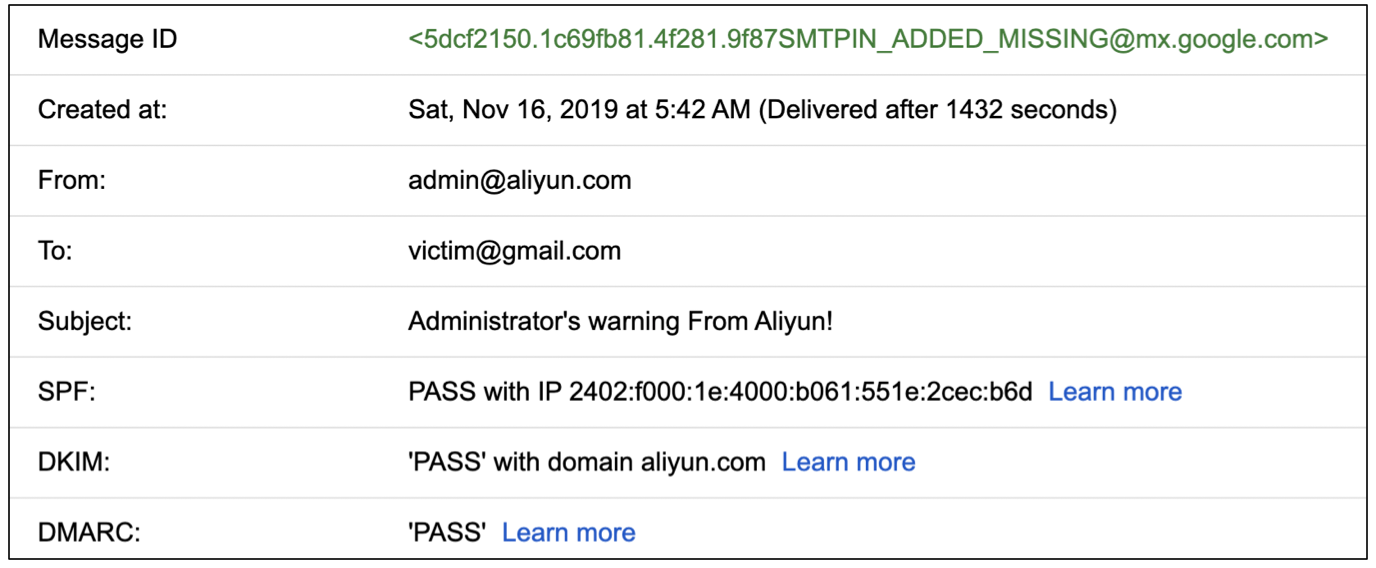}           
\end{minipage}}
\caption{A spoofing example to impersonate $admin@aliyun.com$ via Gmail.}    
\label{fig:example_attack}                                                   
\end{figure}

Figure \ref{fig:example_attack} shows an example of a successful email sender spoofing attack using the direct MTA attack and forward MTA attack models. The attack details are described in Section \ref{combined_attacks}.
All the three email security protocols give "pass" verification results to the spoofing email. Furthermore, the receiver's MUA does not display any security alerts. The victim could hardly recognize any traces of attack from such a seemingly authentic spoofing email. \added[id=skw]{Therefore, it is challenging to identify whether such an email is spoofing, even for people with asenior technical background.}

\subsection{Experimental Target Selection}
We systematically analyze 30 email services, including the most popular free public email services, enterprise-level email services and self-hosted ones.
%Despite our efforts to obtain all public email services analyzed by Hu et al.\cite{hu2018end}, such as Gmail, Outlook, Yahoo, QQ Mail. we have to exclude some email services not opening for registration (e.g., gmx.com, sapo.pt) or  without valid SMTP services (e.g., tutanota.com, protonmail.com). 
Our testing targets include the public email services that have been measured by Hu et al.~\cite{hu2018end}, except for the ones that can neither be registered in China (e.g., gmx.com and sapo.pt) nor have valid SMTP services (e.g., tutanota.com and protonmail.com).

In total, we select 22 popular emails services that have more than 1 billion users.
We believe their security issues can expose a wide range of common users to threats. 
Besides, we also select 5 popular enterprise email services, including Office 365, Alibaba Cloud and Coremail, to test the threat effect on the institutional users.
As for the self-hosted email systems, we build, deploy and maintain 3 famous email systems (i.e., Zimbra, EwoMail, Roundcube).

Further, we test our attacks against 23 widely-used email clients in different desktop and mobile operating systems to evaluate the impact on the UI rendering implementation.

\subsection{Experiment Methodology\label{methodology}}
This work aims to cover all possible verification issues throughout the email delivery process. 
Hence, we conduct a five-step empirical security analysis:

First, we systematically analyze the email specifications. 
In terms of syntax, we extract the ABNF rules~\cite{crocker1997augmented}, focusing on headers (e.g., \texttt{Mail From/From/Helo/Sender} headers) related to authentication. We also pay attention to semantics, particularly the identity verification of emails at each stage in the RFCs. 
Second, we collect legitimate email samples and generate the test samples with authentication-related headers based on the ABNF grammar~\cite{gruber2013extraction}. 
Since common email services usually refuse to handle emails with highly deformed headers, we specify certain header values for our empirical experiment purposes. 
For example, we limit the value of \texttt{domain} to several famous email domain names (e.g., gmail.com, icloud.com). 
Third, we introduce the common mutation methods in protocol fuzzing~\cite{pereyda2017boofuzz}, such as header repeating, inserting spaces, inserting Unicode characters, header encoding, and case variation.
Fourth, we use the generated samples to test the security verification logic of the target email system in four stages. 
Finally, we analyze and summarize the adversarial techniques that make email sender spoofing successful in practice.

\input{Tables/experiment_result.tex}

\subsection{Experiment Setup\label{experiment_setup}}

In this work, we aim to summarize the potential email spoofing methods against the tested email services.
Thus, we try to find out all verification issues from the four stages of the email transmission process mentioned in Section~\ref{sec:background}.
Below, we first introduce the successful attacks from each stage separately.
Then, we discuss our efforts to minimize the measurement bias and avoid ethical problems.

\noindent \textbf{The Successful Attacks. }
We consider an email spoofing attack successful if either of the following four conditions is satisfied.
(1) In the email sending authentication stage, an attacker can modify the identifiers (e.g., \texttt{Auth username}/ \texttt{MAIL From}/ \texttt{From}) arbitrarily.
(2) In the email receiving verification stage, the receiver's MTA gives a "none/pass" verification result even if the spoofed domain name has already deployed strict SPF/DKIM/DMARC policies.
Since the verification results are not always shown in the email headers, we can infer the result by checking whether the email has entered the inbox as an alternative.
Besides, we consider an attack failed if our spoofing email is dropped into the spam box, which means the receiver's MTA has detected the spoofing and taken defensive measures.
To avoid accidental cases, we repeat each attack three times, ensuring that the spoofing email has actually penetrated the security protocols.
Only the attacks that work all three times are regarded as successful attacks.  
(3) \replaced[id=skw]{In the email forwarding stage, the forwarder gives a higher security endorsement to the forwarded email. 
Additionally, an attack is also considered successful if the attacker can freely configure forwarded emails to any accounts without any authentication verification.}{In the email forwarding stage, the forwarder gives a higher security endorsement to the forwarded email in its verification stage.}
(4) In the email UI rendering stage, the displayed email address is inconsistent with the real one. 
In this stage, we use \texttt{APPEND} function of the IMAP\cite{dhamankar2004imap} protocol to deliver the spoofing emails into the inbox, since we only need to check the UI rendering results rather than bypass the spam engine.
Finally, we collect information and analyze the results depend on the webmail and email clients on the UI level. 
% All email services in our experiments have webmail services.

\noindent \textbf{Minimize the Measurement Bias.}
First, to exclude the influence of the spam detection, we select the legitimate, benign and desensitized email samples provided by our industrial partner, a famous email provider, as the contents of our spoofing emails.
These emails' content is legal and harmless and can not be judged as spam.
Second, all spoofing emails are sent from 15 IP addresses located in different regions with an interval of 10 minutes.
Furthermore, we deploy MX/TXT/PTR records for the attacker's domain names and IP addresses. 
Third, to test how the receiver's MTA handles email with "fail" SPF/DMARC verification results, we reproduce the spoofing experiments in Hu's paper\cite{hu2018end} on our target 30 email services. 
We find that 23 of them reject the emails with  "fail" SPF/DMARC verification results. 
The remaining ones mark them as spams.
Besides, the results show that most of the vulnerabilities pointed in Hu's paper\cite{hu2018end} have been fixed in the past two years.

\noindent \textbf{Ethics.} 
We have taken active steps to ensure research ethics. 
Our measurement work only uses dedicated email accounts owned by ourselves. 
No real users are affected by our experiments.  
We have also carefully controlled the message sending rate with intervals over 10 minutes to minimize the impact on the target email services.

\subsection{Experiment Results \label{results}}

This work organizes all testing results in Table~\ref{tab:experiment_results1} and Table~\ref{tab:client_results1} to provide a general picture of the experiment results for sender spoofing attacks.
The details of each attack and spoofing results are discussed in Section ~\ref{email_attacks}.
We summarize our experiment findings as follows.

First, we measured the deployment and verification of email security protocols by these email services.
All email services deploy the SPF protocol on the sender's side, while only 23 services deploy all of the three protocols. 
Surprisingly, all email services run the SPF, DKIM and DMARC detection on the receiver's side.
However, only 12 services perform the sender inconsistency checks. 
Second, all target email services and email clients are vulnerable to certain types of attacks. 
Finally, combined attacks allow attackers to forge 
%more authentic spoofing emails. 
spoofing email which looks more authentic.

\input{Tables/clients.tex}

%% file: Tables/experiment_result.tex
\begin{table*}[!t]
\centering
\small
\caption{Sender spoofing experiment results on 30 target email services.}
\label{tab:experiment_results1}
\begin{threeparttable}
\setlength{\tabcolsep}{1.5mm}{
\begin{tabular}{c|ccc|c|c|c|c|c}
\toprule
%\cline{2-2}
\multicolumn{1}{c|}{\multirow{2}{*}{\textbf{Email Services}}} &\multicolumn{3}{c|}{\textbf{Protocols Deployment}} & \textbf{UI Protections}& \multicolumn{4}{c}{\textbf{Weaknesses in Four Stages of Email Flows}}                                                       \\
%\midrule %表中直线
\multicolumn{1}{c|}{}             & {SPF}          & {DKIM}        & {DMARC}     &SIC &\multicolumn{1}{c}{Sending} & \multicolumn{1}{c}{Receiving}    &\multicolumn{1}{c}{Forwarding} &\multicolumn{1}{c}{UI Rendering}                    \\
\midrule %表中直线
\rowcolor[HTML]{EFEFEF} Gmail.com & \checkmark & \checkmark & \checkmark & \checkmark &      		&  A$_{6}$      		&        	&  A$_{12}$       			\\
									Zoho.com             & \checkmark            & \checkmark           & \checkmark          & \checkmark &A$_2$      	& A$_4$       		&    A$_{11}$    	   	&A$_{13}$             \\
\rowcolor[HTML]{EFEFEF}				iCloud.com           & \checkmark            & \checkmark           & \checkmark          	&  & A$_2$       	&   A$_4$, A$_7$     		& A$_{9}$        	& A$_{12}$     				\\
									Outlook.com          & \checkmark            & \checkmark           & \checkmark      	&     & A$_2$       		& A$_7$       		&  A$_{9}$      	& A$_{14}$        \\
\rowcolor[HTML]{EFEFEF}				Mail.ru          	 & \checkmark            & \checkmark           & \checkmark          & 	&        		&    A$_4$    		&        	& A$_{12}$      \\
									Yahoo.com            & \checkmark            & \checkmark           & \checkmark    &        & A$_2$      & A$_3$,	 A$_7$ & A$_{10}$       		   	& A$_{14}$       \\
\rowcolor[HTML]{EFEFEF}				QQ.com          		 & \checkmark            & \checkmark           & \checkmark & \checkmark         & A$_2$      & A$_5$        		&        		   	&   A$_{13}$, A$_{14}$ \\
									139.com         		 & \checkmark            &            			& \checkmark &    \checkmark   &      		  &   A$_4$    		&                 		   	& A$_{13}$     \\
\rowcolor[HTML]{EFEFEF}				Sohu.com        		 & \checkmark            &            			&         &  		  & A$_2$      &  A$_4$, A$_5$        	&  A$_{9}$  		   	& A$_{13}$     \\
									Sina.com        		 & \checkmark            &            			&       &    		  & A$_2$      & A$_3$, A$_4$, A$_5$, A$_8$       		&        		   	& A$_{13}$, A$_{14}$    \\
\rowcolor[HTML]{EFEFEF}				Tom.com         		 & \checkmark            & \checkmark           & \checkmark        &  & A$_2$       		&        		& A$_{9}$       	&        \\
									Yeah.com        		 & \checkmark            & \checkmark           & \checkmark          & \checkmark & A$_2$      & A$_3$, A$_4$, A$_5$, A$_7$, A$_8$       		& A$_{9}$       		   	& A$_{12}$, A$_{13}$, A$_{14}$  \\
\rowcolor[HTML]{EFEFEF}				126.com         		 & \checkmark            & \checkmark           & \checkmark          & \checkmark & A$_2$       	    & A$_3$, A$_4$, A$_5$, A$_8$        	& A$_{9}$       	& A$_{12}$, A$_{13}$, A$_{14}$       \\
									163.com              & \checkmark            & \checkmark           & \checkmark          & \checkmark & A$_2$      & A$_3$, A$_4$, A$_5$, A$_7$, A$_8$    	    	&   A$_{9}$     		   	& A$_{12}$, A$_{13}$, A$_{14}$      \\
\rowcolor[HTML]{EFEFEF}				Aol.com         	     & \checkmark            & \checkmark           & \checkmark        &  & A$_2$      & A$_5$, A$_7$     &        			& A$_{14}$       \\
									Yandex.com      		 & \checkmark            & \checkmark           & \checkmark       &   &       & A$_3$, A$_4$,A$_6$, A$_7$, A$_8$    & A$_{9}$       	& A$_{14}$      \\
%\rowcolor[HTML]{EFEFEF}				Inbox.lv         	 & \checkmark            & \checkmark           & \checkmark        &  &       &     &        			&       \\
%									Runbox.com      		 & \checkmark            & \checkmark           & \checkmark          & -       		& -       		& -       			& -    \\
%\rowcolor[HTML]{EFEFEF}				Rediffmail.com       & \checkmark            & \checkmark           & \checkmark          & -       		& -       		& -       			& -      \\
\rowcolor[HTML]{EFEFEF}				Rambler.ru       	 & \checkmark            & \checkmark           & \checkmark       &   & A$_2$       		& A$_3$    &        			&          \\
									Naver.com       		 & \checkmark            & \checkmark           & \checkmark      &    & A$_2$      & A$_4$, A$_5$, A$_8$      &        			&        \\
\rowcolor[HTML]{EFEFEF}				21cn.com        		 & \checkmark            &            			&         &  		  & A$_2$      & A$_4$, A$_5$     & A$_{9}$        	&         \\
									Onet.pl          	 & \checkmark            &            			&           		&  & A$_2$      & A$_4$, A$_5$    &         &           \\
\rowcolor[HTML]{EFEFEF}				Cock.li          	 & \checkmark            & \checkmark           &     &      		  & A$_2$      & A$_3$, A$_4$    &        			& A$_{13}$, A$_{12}$        \\
%\rowcolor[HTML]{EFEFEF}				Freemail.hu      	 & \checkmark            & \checkmark           &           		  & -      		 	& -       		& -       			& -     \\
%									Freenet.de       	 & \checkmark            &            			& \checkmark          & -       		& -       		& -       			& -       \\
									Daum.net        	 	 & \checkmark            &            			& \checkmark          &   & & A$_5$           		&        			&      \\
\rowcolor[HTML]{EFEFEF}				Hushmail.com         & \checkmark            & \checkmark           & \checkmark        &  &       & A$_3$, A$_4$, A$_8$    &         & A$_{12}$         \\
									Exmail.qq.com        & \checkmark            & \checkmark           & \checkmark          & \checkmark & A$_2$      & A$_5$     &         &  A$_{14}$       \\
\rowcolor[HTML]{EFEFEF}				Coremail.com         & \checkmark            & \checkmark           & \checkmark        & \checkmark  & A$_2$      & A$_8$    & A$_{9}$        &   \\
									Office 365       	 & \checkmark            & \checkmark           & \checkmark       & \checkmark   & A$_2$      & A$_4$    & A$_{9}$, A$_{10}$,A$_{11}$        & A$_{14}$    \\
\rowcolor[HTML]{EFEFEF}				Alibaba Cloud    		 	 & \checkmark            & \checkmark           & \checkmark         & \checkmark & A$_2$      & A$_3$, A$_4$, A$_5$, A$_8$   & A$_{10}$         &  A$_{13}$      \\

									Zimbra           	 & \checkmark            & \checkmark           & \checkmark          & \checkmark & A$_1$, A$_2$      & A$_3$, A$_5$, A$_8$    & A$_{9}$         & A$_{12}$, A$_{13}$        \\
\rowcolor[HTML]{EFEFEF}				EwoMail          	 & \checkmark            & \checkmark           & \checkmark       &   & A$_2$      & A$_3$, A$_4$, A$_8$    &         &  A$_{13}$      \\
									Roundcube        	 & \checkmark            & \checkmark           & \checkmark         &  & A$_1$, A$_2$      & A$_3$, A$_4$, A$_8$    &         & A$_{12}$      \\
\midrule %表中直线
%							\textbf{Percentage}      & 100\%     		 & 75.9\%     			 & 75.9\%    			& \% 			  & \% 			& \% 	    & \% 		\\
%\midrule %表中直线
\end{tabular}}
%\leftline{-\; : No related features, such as SMTP login and automatic forwarding.}
%\leftline{x\,\,: The sender does not deploy the security protocol or is no vulnerabilities.}
%\leftline{\checkmark: The sender deploys the security protocol or is vulnerabilities through the SMTP attack.}
\begin{tablenotes}
\item[1] The subscript identifies the specific attack (e.g., A$_8$ identifies the encoding based attack discussed in ~\ref{A8}).
\item[2] The abbreviation SIC stands for the receiver's sender inconsistency checks, an email notification custom deployed by providers, described in the background ~\ref{SIC}.
\item[3] The cases with \checkmark\, mean that the domain name deploys with the relevant email security protocol or perform the sender inconsistency checks.
\end{tablenotes}
\end{threeparttable}
\end{table*}

%% file: Tables/clients.tex
% \usepackage{multirow}
\begin{table}[!t]
\caption{Sender spoofing experiment results on 23 target email clients.}
\label{tab:client_results1}
\small
\centering
\begin{threeparttable}
\setlength{\tabcolsep}{1mm}{
%\begin{tabular}{m{1cm}|m{2cm}|m{3cm}m{3cm}m{3cm}m{3cm}m{3cm}m{3cm}m{3cm}m{3cm}}
\begin{tabular}{c|c|c|c}
\toprule

OS                       & Clients      & \tabincell{c}{SIC} & \tabincell{c}{Weaknesses}  \\
\midrule %表中直线
\multirow{5}{*}{\rotatebox{0}{Windows}} & Foxmail     & \checkmark   &    A$_6$, A$_7$, A$_{13}$, A$_{14}$                 \\
                         & Outlook     & \checkmark  &  A$_6$, A$_{13}$                   \\
                         & eM Client   & \checkmark  &    A$_6$, A$_{12}$                 \\
                         & Thunderbird & & A$_6$, A$_{13}$, A$_{14}$                    \\
                         & Windows Mail& & A$_6$, A$_7$, A$_{13}$, A$_{14}$                     \\
\midrule %表中直线
\multirow{5}{*}{\rotatebox{0}{MacOS}}     & Foxmail     &   & A$_6$, A$_{13}$                    \\
                         & Outlook     & \checkmark  &  A$_6$, A$_{13}$                \\
                         & eM Client  & \checkmark  &      A$_6$, A$_7$, A$_{12}$, A$_{13}$, A$_{14}$               \\
                         & Thunderbird & &  A$_6$, A$_{13}$, A$_{14}$                   \\
                         & Apple Mail   & & A$_6$, A$_{13}$, A$_{14}$                    \\
\midrule %表中直线
\multirow{5}{*}{\rotatebox{0}{Linux}}   & Thunderbird & &  A$_6$, A$_{13}$                   \\
                         & Mailspring  & &  A$_6$, A$_{13}$, A$_{14}$                   \\
                         & Claws Mail  & &  A$_6$, A$_{14}$                   \\
                         & Evolution   & &  A$_6$, A$_{13}$, A$_{14}$                   \\
                         & Sylpheed    & &   A$_6$, A$_{13}$, A$_{14}$                  \\
\midrule %表中直线
\multirow{3}{*}{\rotatebox{0}{Android}} & Gmail       & &  A$_6$, A$_{13}$                   \\
                         & QQ Mail     & \checkmark  & A$_6$, A$_{13}$, A$_{14}$                    \\
                         & NetEase Mail &  &  A$_6$, A$_{12}$, A$_{13}$ \\
                         & Outlook & \checkmark  &  A$_6$, A$_{13}$ \\
\midrule %表中直线
\multirow{3}{*}{\rotatebox{0}{iOS}}     & Mail.app    & & A$_6$, A$_7$, A$_{13}$, A$_{14}$                     \\
                         & QQ Mail     & \checkmark &   A$_6$, A$_{13}$                  \\  
                         & NetEase Mail &  & A$_6$, A$_{12}$, A$_{13}$ \\  
                         & Outlook & \checkmark &  A$_6$, A$_{13}$ \\      
\bottomrule
\end{tabular}
}
\begin{tablenotes}
\item[1] The subscript identifies the specific attack.
\item[2] The SIC stands for the sender inconsistency checks.
\item[3] The cases with \checkmark\, mean that the email client performs the sender inconsistency checks.
\item[4] Since email clients do not involve verification of the mail protocol, we only tested attacks (i.e., A$_6$, A$_7$, A$_{12}$, A$_{13}$, A$_{14}$) related to email UI rendering.
\end{tablenotes}
\end{threeparttable}
\end{table}

%% file: 4_attacks.tex
\section{Email Sender Spoofing Attacks \label{email_attacks}}

This section describes the various techniques employed in email spoofing attacks. 
We divide the attacks into four categories, corresponding to the four authentication stages in the email delivery process.

\subsection{Attacks in Email Sending Authentication}
Email sending veriﬁcation is a necessary step to ensure email authenticity.
Attacks in email sending authentication can abuse the IP reputation of a well-known email service. 
They can even bypass all the verification of SPF/DKIM/DMARC protocols, which poses a significant threat to the email security ecosystem. These attacks are mainly used in the shared attack model (Model \textcircled{a}).

As mentioned in Section \ref{multiple_identity}, there are three sender identifiers in email sending process: (1) \texttt{Auth username}; (2) \texttt{Mail From}; (3) \texttt{From}. 
\replaced[id=skw]{ An attack is considered successful while it can arbitrarily control these identifiers during email sending authentication process.
}{
If an attacker can arbitrarily control these identifiers during the email sending authentication process, the attack is considered successful.} 

\begin{figure}[t]
\centering                                                          
\subfigure[Attack with different auth username and Mail From header]{
\label{fig:sender_attack1}                    
\begin{minipage}{8cm}
\centering                                               
\includegraphics[width=2in]{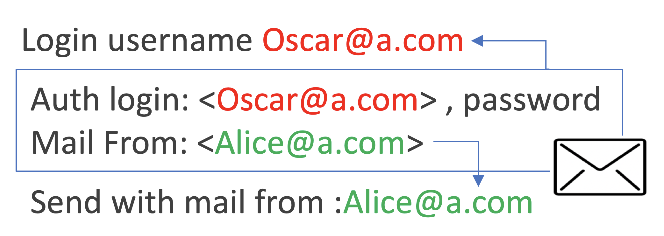}               
\end{minipage}
} 
\quad
\subfigure[Attack with different Mail From and From headers]{
\label{fig:sender_attack2}    
\begin{minipage}{8cm}
\centering         
\includegraphics[width=2in]{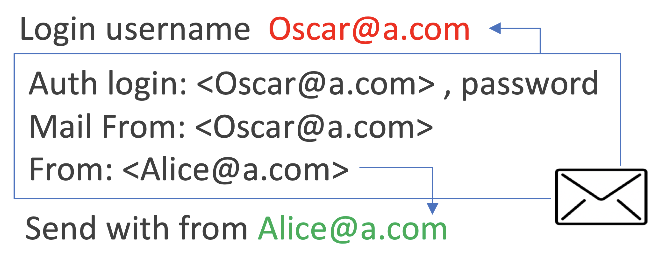}           
\end{minipage}}
\caption{Two attacks of bypassing sending service's verification.}                                                    
\end{figure}

% \myparagraph{}
\noindent \textbf{The Inconsistency between Auth username and Mail From headers (A$_1$)\label{A$_1$}.}
As shown in Figure \ref{fig:sender_attack1}, an attacker can pretend to be any user under the current domain name to send a spoofing email whose \texttt{Auth username} (Oscar@a.com) and \texttt{Mail From} (Alice@a.com) are inconsistent during email sending authentication. 
SMTP protocol does not provide any built-in security features to guarantee the consistency of \texttt{auth username} and \texttt{Mail From} header.
Therefore, this type of protection depends only on the software implementation of the email developer. 

In our spoofing experiments, most email services have noticed such problems and prohibited users from sending emails inconsistent with their original identity. However, this type of problem still appears in some well-known corporate email software (i.e., Zimbra, EwoMail). These two email services are vulnerable under default security configuration. Email administrators need to upgrade their security configurations to prevent such problems manually.

% \myparagraph{}
\noindent \textbf{The Inconsistency between \texttt{Mail From} and \texttt{From} headers (A$_2$)\label{A2}.}
An attacker can send a spoofing email with different \texttt{Mail From} and \texttt{From} headers. 
Figure \ref{fig:sender_attack2} shows this type of attack.
Although some users are allowed to use email aliases to send emails with a different \texttt{From} header, no user should be allowed to freely modify the \texttt{From} header to any value (e.g., admin@a.com) to prevent attacks. 
The \texttt{From} header should only be allowed to be set within limited legal values. 
Many prevalent email services (e.g., Outlook, Sina, QQ Mail) and most third-party email clients (e.g., Foxmail, Apple Mail) only display the \texttt{From} header, not the \texttt{Mail From} header.
For these emails which have different \texttt{Mail From} and \texttt{From} headers, the victim cannot even see any security alerts on the MUA.

Similar inconsistency also exists between the \texttt{RCPT To} and \texttt{To} headers. 
In the real world, there are some scenes that cause the inconsistency, such as email forwarding and Bcc. 
However, this kind of flexibility increases attack surfaces and introduces new security risks. For example, an attacker can send an email to a victim, even if the email's \texttt{To} header is not the address of the victim. 
In this case, an attacker can further use this method to obtain a spoofing email with a DKIM signature that normally could not be obtained, which is helpful for further attacks. 
This technique might not be effective when used alone, but it can often achieve excellent spoofing results when combined with other attack techniques.

14 email services are vulnerable to this type of attack in our experiments. 
In addition, we also found that some email services (e.g., Outlook, Zoho, AOL, Yahoo) have realized these risks and have implemented corresponding security restrictions. 
They refused to send emails with inconsistent \texttt{Mail From} and \texttt{From} headers during SMTP sending process. 
However, these defenses can still be bypassed by two types of attacks (i.e., A$_4$, A$_5$). 
For example, we can send a spoofing email with the \texttt{Mail From} header as \texttt{<Oscar@a.com>} and the \texttt{From} header as \texttt{<Alice@a.com, Oscar@a.com>} in Yahoo which introduces another source of ambiguity and eventually bypasses email protocol verification.
Therefore, it is still possible to send such spoofing emails, even if the sender has deployed relevant security measures.

\subsection{Attacks in Email Receiving Verification}
SPF, DKIM and DMARC are the prevalent mechanisms used to counter email spoofing attacks. 
If an attacker can bypass these protocols, it can also pose a serious security threat to email security ecosystem. 
There are three attack models to launch this type of attack: shared MTA attack, direct MTA attack, and forward MTA attack.  
An attack is successful while the receiver's MTA incorrectly gets a 'none/pass' verification result.

\begin{figure}[t]
\centering
\includegraphics[width=2.3in]{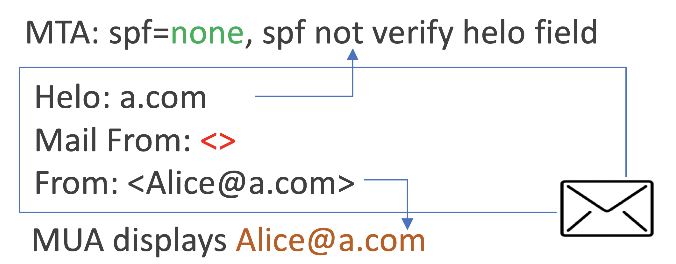}
\caption{Empty \texttt{Mail From} attack bypassing the SPF verification.}
\label{fig:empty_mail_attack}
\end{figure}

% \myparagraph{}
\noindent \textbf{Empty Mail From Attack (A$_3$)\label{A3}.}
RFC 5321\cite{klensin2008simple} explicitly describes that an empty \texttt{Mail From} is allowed, which is mainly used to prevent bounce loop-back and allow some special message. 
However, this feature can also be abused to launch email spoofing attacks. 
As shown in Figure \ref{fig:empty_mail_attack}, an attacker can send an email with an empty \texttt{Mail From} header, and the \texttt{From} header fabricates Alice's identity (Alice@a.com). 

The SPF protocol\cite{kitterman2014rfc} stipulates that the receiver's MTA must complete the SPF verification based on the \texttt{Helo} field if the \texttt{Mail From} header is empty.
However, the abuse of the \texttt{Helo} field in real life make some email services disobey the standard and take a more loose approach of verification.
Thus, when the recipient deals with those emails, they can not complete SPF verification based on the \texttt{Helo} field, but directly return "none". 
This type of error allows an attacker to bypass the SPF protection. 
As a result, an attacker can change the SPF result of this attack from "fail" to "none".

13 email services (e.g., Yahoo, Yeah, 126, Aol) are vulnerable to this type of attacks. 
Fortunately, there are already 17 email services that have fixed such security issues, 5 of which (e.g., Zoho.com, iCloud.com, exmail.qq.com) drops such emails into spam.

\begin{figure}[t]
\centering
\vspace{-0.1cm}
\subfigure[Ordinary multiple \texttt{From} attack.]{
    \begin{minipage}[t]{0.5\linewidth}
        \centering
        \includegraphics[width=1.65in]{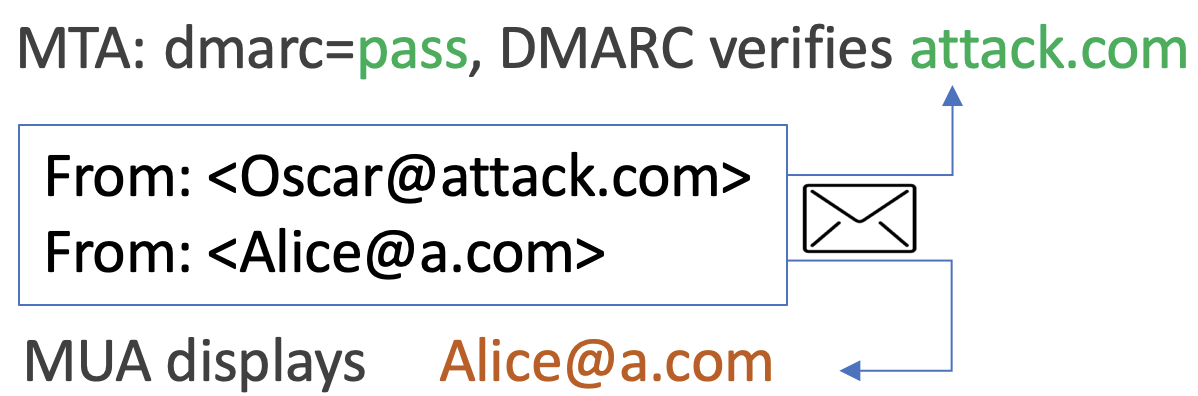}\\
        \vspace{0.02cm}
        \label{fig:multiple_from_attack1}
    \end{minipage}%
}%
\subfigure[Multiple \texttt{From} attack with spaces.]{
    \begin{minipage}[t]{0.5\linewidth}
        \centering
        \includegraphics[width=1.65in]{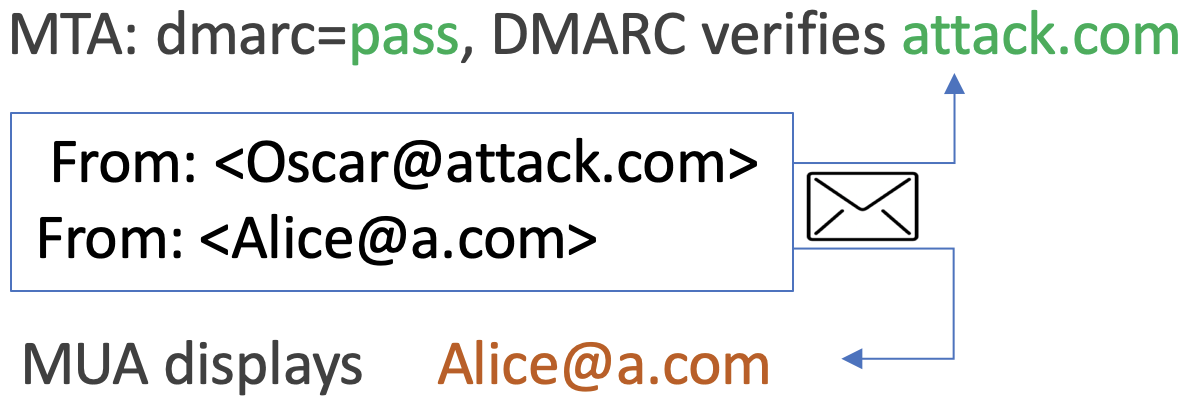}\\
        \vspace{0.02cm}
        \label{fig:multiple_from_attack2}
    \end{minipage}%
}%
\quad
\subfigure[Multiple \texttt{From} attack with case \quad variation.]{
    \begin{minipage}[t]{0.5\linewidth}
        \centering
        \includegraphics[width=1.65in]{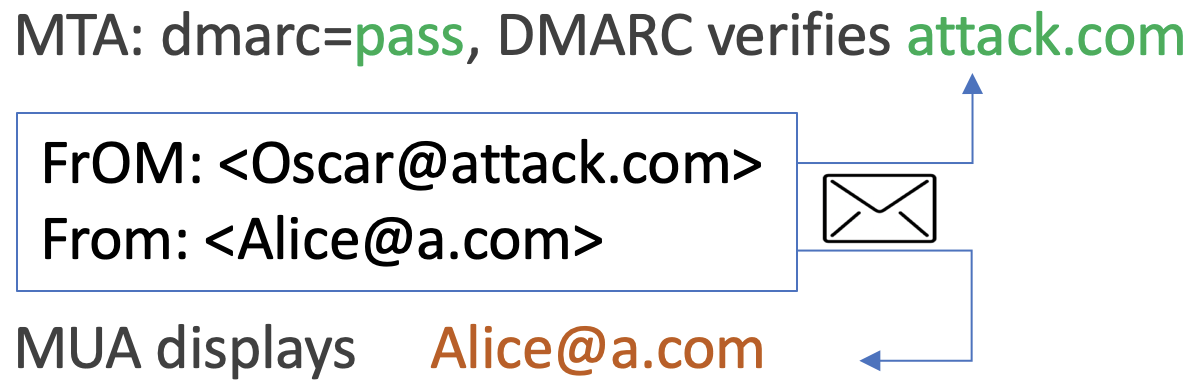}\\
        \vspace{0.02cm}
        \label{fig:multiple_from_attack3}
    \end{minipage}%
}%
\subfigure[Multiple \texttt{From} attack with invisible characters.]{
    \begin{minipage}[t]{0.5\linewidth}
        \centering
        \includegraphics[width=1.65in]{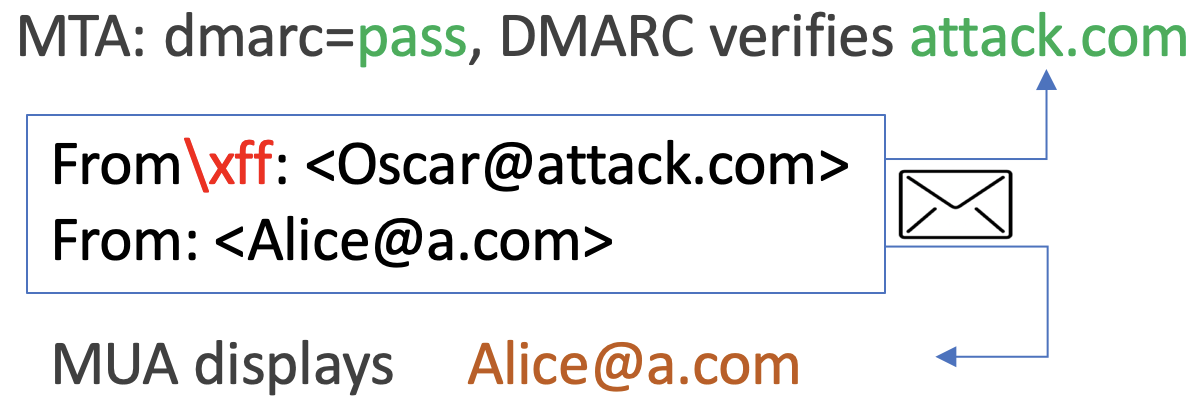}\\
        \vspace{0.02cm}
        \label{fig:multiple_from_attack4}
    \end{minipage}%
}%

\centering
\caption{Multiple \texttt{From} attacks to make DMARC verify Oscar@attack.com while the MUA displays Alice@a.com.}
\vspace{-0.1cm}
\label{fig:multiple_from_attack}
\end{figure}

% \myparagraph{}
\noindent \textbf{Multiple From Headers (A$_4$)\label{attack:A$_4$}.}
Inspired by the work of Chen~et~al.\cite{chen2016host}, we also utilize multiple headers techniques in email spoofing attacks. 
Compared with Chen's work, we have more distortions from the \texttt{From} header, such as adding spaces before and after the \texttt{From}, case conversion, and inserting non-printable characters. 
As shown in Figure \ref{fig:multiple_from_attack}, an attacker can construct multiple \texttt{From} headers to bypass security policies.
RFC 5322\cite{resnick2008rfc} indicates that emails with multiple \texttt{From} fields are typically rejected.
However, there are still some email services that fail to follow the protocol and accept emails with multiple \texttt{From} headers.
It can introduce inconsistencies in the email receiving verification stage, which could lead to additional security risks. 
Figure \ref{fig:multiple_from_attack3} shows an example that the displayed sender address is Alice@a.com, while the receiver's MTA may use Oscar@attack.com for the DMARC verification . 

Only 4 mail services (i.e., Gmail, Yahoo, Tom, Aol) reject emails with multiple \texttt{From} headers, and 19 mail services are affected by this type of attacks. 
Most tested email services tend to display the first \texttt{From} header on the webmail, while 6 services (e.g., iCloud, Yandex, Alibaba Cloud) choose to display the last \texttt{From} header. 
Besides, 7 vendors have made specific security regulations against such attacks, such as showing two \texttt{From} addresses on the webmail simultaneously (e.g., QQ Mail, Coremail)  or dropping such emails into the spam folder (e.g., Outlook, rambler.ru). 

\begin{figure}[t]
\centering
\subfigure[Ordinary multiple address attack.]{
    \begin{minipage}[t]{0.5\linewidth}
        \centering
        \includegraphics[width=1.7in]{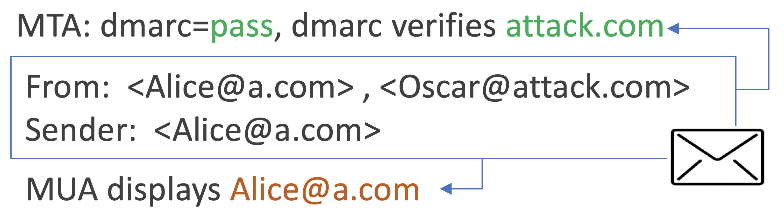}\\
        \vspace{0.02cm}
		\label{fig:multiple_address_attack1}
    \end{minipage}%
    \label{fig:}
    
}%
\subfigure[Multiple address attack with null address.]{
    \begin{minipage}[t]{0.5\linewidth}
        \centering
        \includegraphics[width=1.7in]{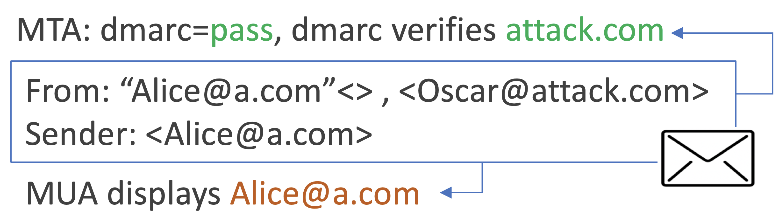}\\
        \vspace{0.02cm}
		\label{fig:multiple_address_attack2}
    \end{minipage}%
}%
\quad
\subfigure[Multiple address attack with semantic characters.]{
    \begin{minipage}[t]{0.5\linewidth}
        \centering
        \includegraphics[width=1.7in]{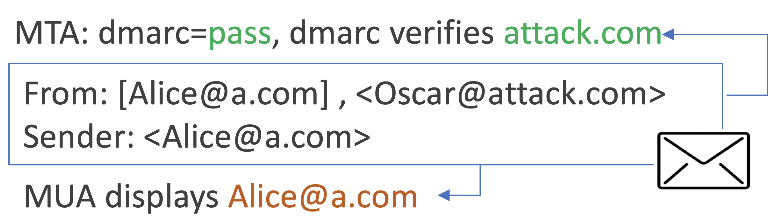}\\
        \vspace{0.02cm}
		\label{fig:multiple_address_attack3}
    \end{minipage}%
}%
\subfigure[Multiple address attack with comments.]{
    \begin{minipage}[t]{0.5\linewidth}
        \centering
        \includegraphics[width=1.7in]{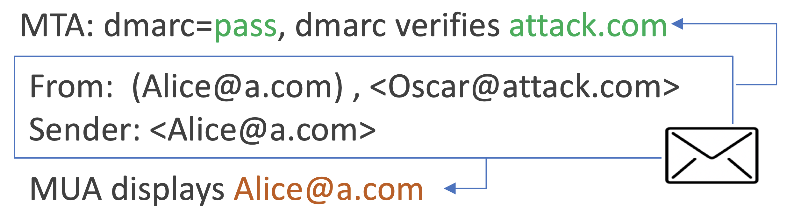}\\
        \vspace{0.02cm}
		\label{fig:multiple_address_attack4}
    \end{minipage}%
}%
\caption{Multiple email addresses attacks to make DMARC verify Oscar@attack.com while MUA displays Alice@a.com.}
\vspace{-0.2cm}
\label{fig:compare_fig}
\end{figure}

\begin{figure*}[htbp]
\centering
 \subfigure[Parsing inconsistency with route portion.]{
    \begin{minipage}[t]{0.33\linewidth}
        \centering
        \includegraphics[width=2.1in]{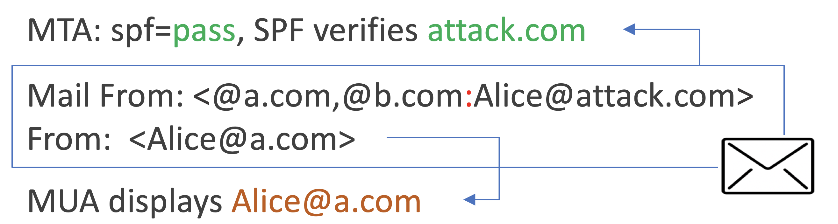}\\
        \vspace{0.02cm}
        \label{fig:parsing_attack1}
    \end{minipage}%
}%
\subfigure[Parsing inconsistency with "null" mailbox-list.]{
    \begin{minipage}[t]{0.33\linewidth}
        \centering
        \includegraphics[width=2.1in]{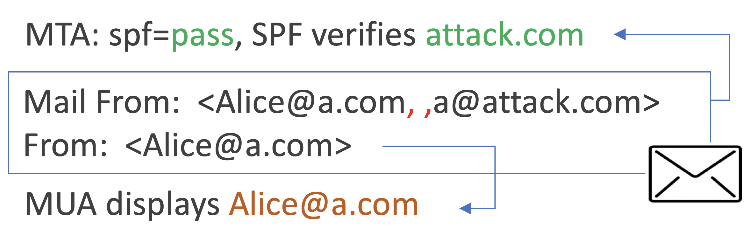}\\
        \vspace{0.02cm}
        \label{fig:parsing_attack2}
    \end{minipage}%
}%
\subfigure[Parsing inconsistency with comment.]{
    \begin{minipage}[t]{0.33\linewidth}
        \centering
        \includegraphics[width=2.1in]{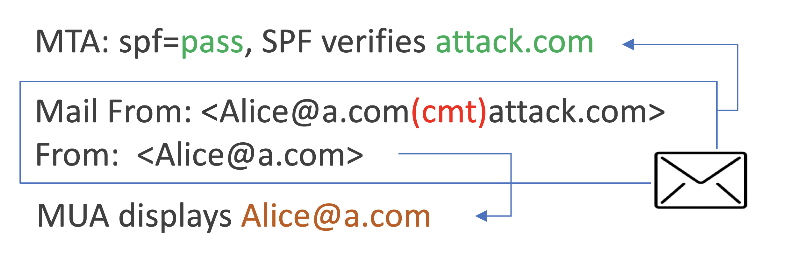}\\
        \vspace{0.02cm}
        \label{fig:parsing_attack3}
    \end{minipage}%
}%
\quad
\subfigure[NUL character truncates string parsing.]{
    \begin{minipage}[t]{0.33\linewidth}
        \centering
        \includegraphics[width=2.1in]{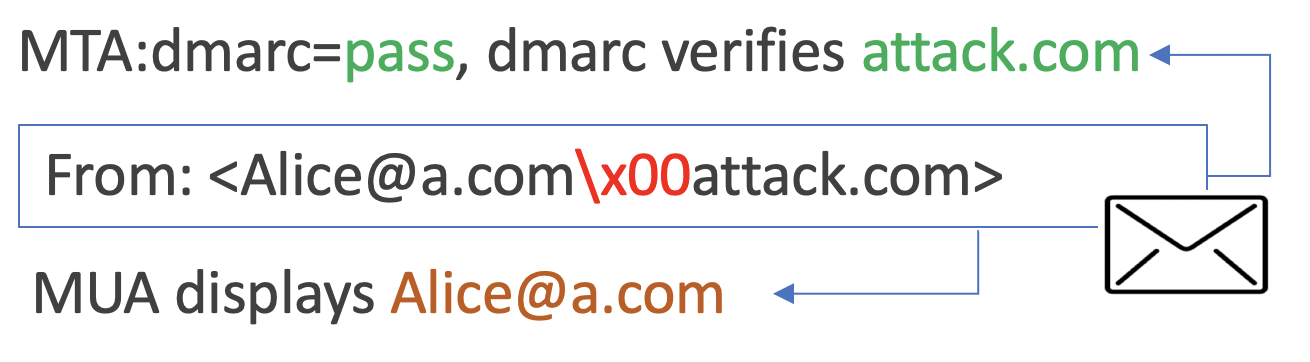}\\
        \vspace{0.02cm}
        \label{fig:terminate1}
    \end{minipage}%
    
}%
\subfigure[Invisible unicode characters truncate string parsing.]{
    \begin{minipage}[t]{0.33\linewidth}
        \centering
        \includegraphics[width=2.1in]{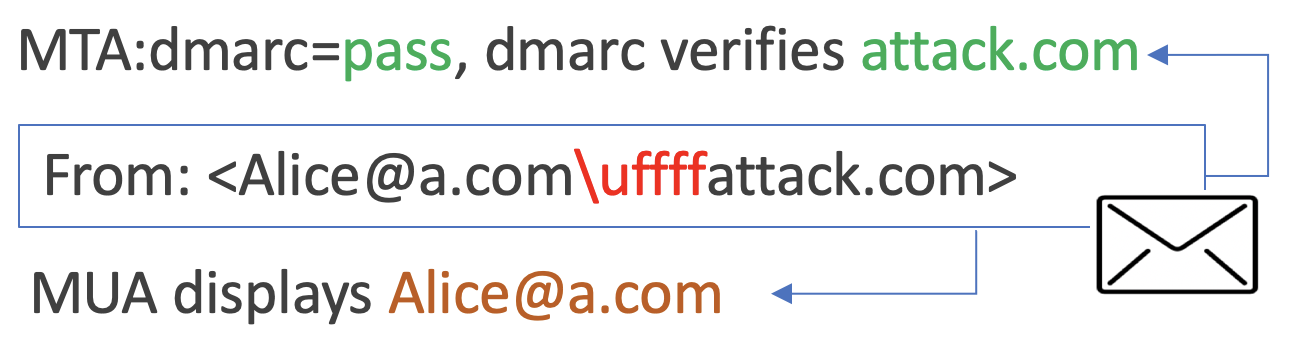}\\
        \vspace{0.02cm}
        \label{fig:terminate2}
    \end{minipage}%
}%
\subfigure[Semantic characters truncate string parsing.]{
    \begin{minipage}[t]{0.33\linewidth}
        \centering
        \includegraphics[width=2.1in]{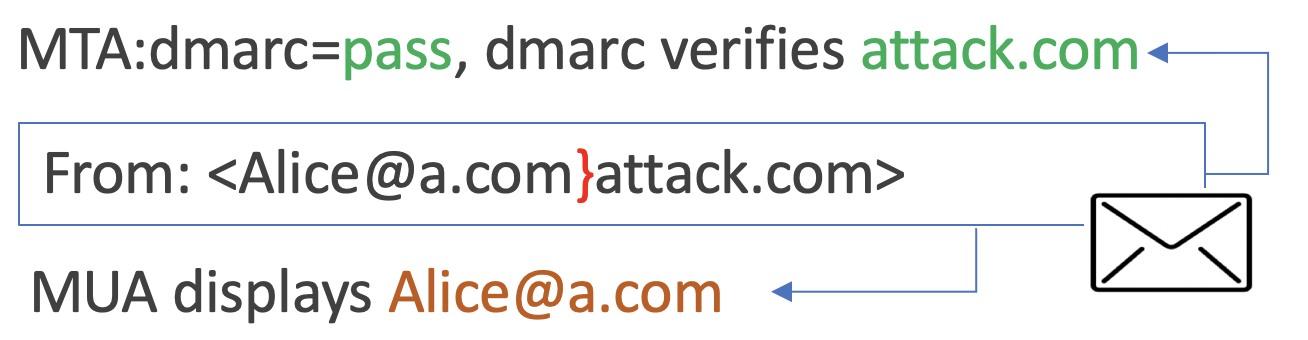}\\
        \vspace{0.02cm}
        \label{fig:terminate3}
    \end{minipage}%
}%

\centering
\caption{Six spoofing examples of bypassing receiving service's verification.}
\vspace{-0.2cm}
\label{fig:spoofing_examples}
\end{figure*}

% \myparagraph{}
\noindent \textbf{Multiple Email Addresses (A$_5$)\label{A5}.}
Using multiple email addresses is also an effective technique to bypass protocol verification. 
Usage of multiple addresses was first proposed in RFC2822\cite{resnick2001rfc2822} and is still explicitly allowed in RFC 5322~\cite{resnick2008rfc}. 
It is suitable for such scenarios: an email with multiple authors is supposed to list all of them in the \texttt{From} header. 
Then, the \texttt{Sender} field is added to mark the actual sender. 
As shown in Figure~\ref{fig:multiple_address_attack1}, an attacker can bypass DMARC verification with multiple email addresses (\texttt{<Alice@a.com>, <Oscar@attack.com>}). 
In addition, we can also make some rule-based mutations to these addresses, such as \texttt{[Alice@a.com], <Oscar@attack.com>}.

15 mail services (e.g., QQ mail, 21cn.com and onet.pl) would still accept such emails. 
Only 4 services (e.g., Gmail and Mail.ru) directly reject those emails, and 5 other services (e.g., zoho.com, tom.com, outlook.com) put them into spam. 
The rest 6 services (e.g., 139.com, cock.li and Roundcube) display all of these addresses, making spoofing emails more difficult to deceive the victim.

% \myparagraph{}
\noindent \textbf{Parsing Inconsistencies Attacks (A$_6$)\label{A6}.}
\texttt{Mail From} and \texttt{From} headers are in rich text with a very complicated grammatical format. 
As a result, it is challenging to parse display names and real addresses correctly. 
These inconsistencies can allow attackers to bypass authentication and spoof their target email clients.

A mailbox address is one of the essential components of these two headers.
First, mailbox addresses were allowed to have a route portion\cite{resnick2001rfc2822} in front of the real sender address when enclosed in "<" and ">". 
Therefore, the mailbox (\texttt{<@a.com, @b.com:admin@c.com>}) is still a legal address. 
Among them, \texttt{@a.com, @b.com} is the route portion, and "admin@c.com" is the real sender's address.
Second, it is allowed to use mailbox-list and address-list\cite{resnick2001rfc2822}, and they can have "null" members, such as \texttt{<a@a.com>, ,<b@b.com>}. 
Third, comment\cite{resnick2008rfc} is a string enclosed in parentheses. They were allowed between the period-separated elements of local-part and domain, such as \texttt{<admin(username)@a.com(domain name)>}. 
Finally, there is an optional display-name\cite{resnick2008rfc} in the \texttt{From} header. It indicates the sender's name, which is displayed for receivers. 
Figure \ref{fig:spoofing_examples} shows three types of attacks based on parsing inconsistencies.

%\myparagraph{Truncated characters(A$_7$)\label{A7}}
Truncated characters are a series of characters that terminate string parsing. 
When parsing and extracting the target domain name from the email headers, truncated characters will end the parsing process. 
Figure \ref{fig:terminate1} shows that the program gets an incomplete domain name (a.com) when parsing the target domain name from the string "admin@a.com$\backslash$x00@attack.com". 
Attackers can use these techniques to bypass the verification of email security protocols.
Overall, this work finds three types of truncated characters in the email string parsing process. 
First, NUL ($\backslash$x00) character can terminate string in the C programming language. 
It has the same effect in the email field.
Second, some invisible Unicode characters (e.g., \texttt{$\backslash$uff00-$\backslash$uffff},\texttt{$\backslash$x81-$\backslash$xff}) can also terminate the string parsing process. 
Third, certain semantic characters, such as "\texttt{[},\texttt{]},\texttt{\{},\texttt{\}},\texttt{$\backslash$t},\texttt{$\backslash$r},\texttt{$\backslash$n},\texttt{;}", can be used to indicate a tokenization point in lexical analysis. 
Meanwhile, these characters also influence the string parsing process.

We found that 13 email services have problems in the UI rendering stage under such attacks. For Gmail and Yandex, we can use these attack techniques to bypass DMARC.

\begin{figure}[t]
\centering
\subfigure[Encoding based attack bypassing DMARC verification.]{
    \begin{minipage}[t]{1\linewidth}
        \centering
        \includegraphics[width=2.3in]{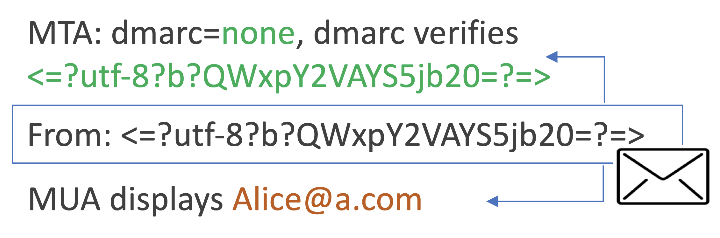}\\
        \vspace{0.02cm}
		\label{fig:encoding_attack1}
    \end{minipage}%

}%
\quad
\subfigure[Combined encoding and truncated attack.]{
    \begin{minipage}[t]{1\linewidth}
        \centering
        \includegraphics[width=2.3in]{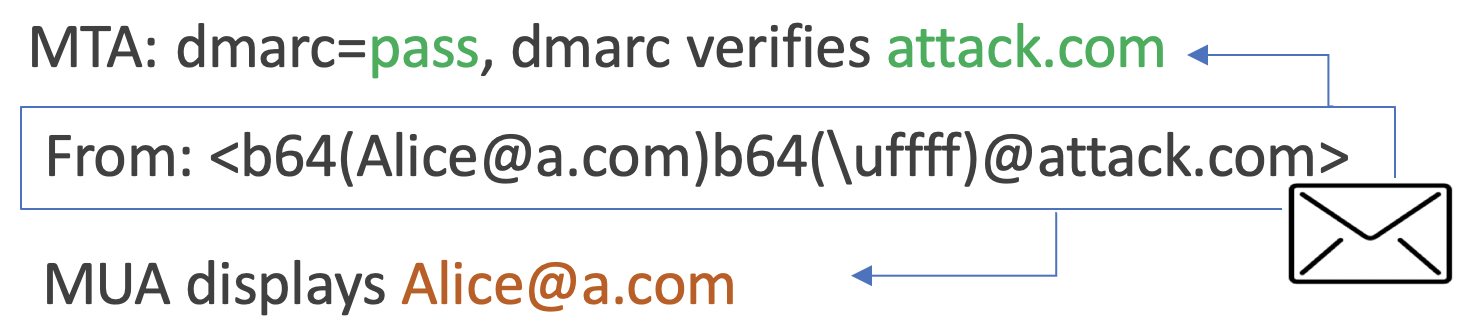}\\
        \vspace{0.02cm}
        \label{fig:encoding_attack2}
    \end{minipage}%
}%

\caption{Two spoofing examples with encoding based attacks.}
\vspace{-0.2cm}
\label{fig:encoding_attacks}
\end{figure}

% \myparagraph{}
\noindent \textbf{Encoding Based Attack (A$_7$)\label{A7}.}
RFC 2045(MIME)\cite{freed1996multipurpose} describes a mechanism denoting textual body parts, which are coded in various character sets. 
The ABNF grammar of these parts is as follows:\texttt{=?charset?encoding?encoded-text?=}. 
The "charset" field specifies the character set associated with the not encoded text; "encoding" field specifies the encoding algorithm, where "b" represents base64 encoding, and "q" represents quoted-printable encoding; "encode-text" field specifies the encoded text.	
Attackers can use these encoded addresses to evade email security protocol verification. 
Figure \ref{fig:encoding_attack1} shows the details such attacks.
For an encoded address, such as \texttt{From: =?utf-8?b?QWxpY2VAYS5jb20=?=}, most email services do not decode the address before verifying the DMARC protocol, thus fail to extract the accurate domain and get a "None" in the following DMARC verification.
However, some email services  display the decoded sender address (\texttt{Alice@a.com}) on the MUA. 
Furthermore, this technique can be combined with truncated strings. 
As shown in the Figure \ref{fig:encoding_attack2}, an attacker can construct the From header as "b64(Alice@a.com>b64($\backslash$uffff)@attack.com". 
%When displayed in an email client, the program might get an incomplete username (i.e., Alice@a.com), nevertheless, it would still use the attacker's domain (attack.com) for DMARC verification.
Email client programs could get incomplete username(i.e., Alice@a.com), but it would still use the attacker's domain (attack.com) for DMARC verification.

7 email services are affected by the vulnerability, including some popular services (e.g., Outlook, Office 365, Yahoo) with more than one billion users.

% \myparagraph{}
\noindent \textbf{The Subdomain Attack (A$_8$)\label{A8}.}
An attacker can send spoofing emails from a non-existent subdomain (no MX record) of well-known email services (e.g., admin@mail.google.com).
Thus, there are no corresponding SPF records. 
The spoofing email only gets a "None" verification result, and the receiver's MTA does not directly reject it.
Although the parent domain (e.g., google.com) deploys strict email policies, attackers can still attack in this way. 
Unfortunately, many companies use sub-domains to send business subscription emails, such as Paypal, Gmail, and Apple. 
As a result, ordinary users tend to trust such emails. 

Unfortunately, RFC 7208\cite{kitterman2014sender} states that the use of wildcard records for publishing SPF records is discouraged. 
And few email administrators configure wildcard SPF records in the real world. 
Besides, the receiver's MTA can usually reject emails from domains without an MX record. 
But RFC 2821\cite{klensin2001rfc} mentions that, when a domain has no MX records, SMTP assumes an A record will suffice, which means any domain name with an A record can be considered a valid email domain. 
In addition, many well-known websites deploy a wildcard DNS A record that makes this type of attack more applicable. 
As a result, it is difficult for the receiver's MTA to determine whether to reject such emails.

Experimental results show that 13 email services are vulnerable to such attacks.
Only one email service (Mail.ru) deploys a wildcard DNS entry for the SPF record in our experiments. 
By default, the DMARC policy set for an organizational domain should apply to any sub-domains, unless a DMARC record has been published for a specific sub-domain. 
However, the experimental results show that our attack is still effective, even if the receiver's MTA conducted a DMARC check.

% \begin{figure}[!t]
% \centering
% \includegraphics[width=6cm]{Figs/8_subdomain_attack.png}
% \caption{The subdomain spoofing attack.}
% \label{fig:subdomain-attack}
% \end{figure}

\subsection{Attacks in Email Forwarding Verification}

This work shows that attackers can abuse the email forwarding service to send spoofing emails that would fail in the shared MTA attack model.
Besides, forwarding service may give the forwarded email a higher security endorsement. 
Both situations are exploitable for attackers to send  spoofing emails.

\begin{figure}[t]
\centering
\includegraphics[width=8cm]{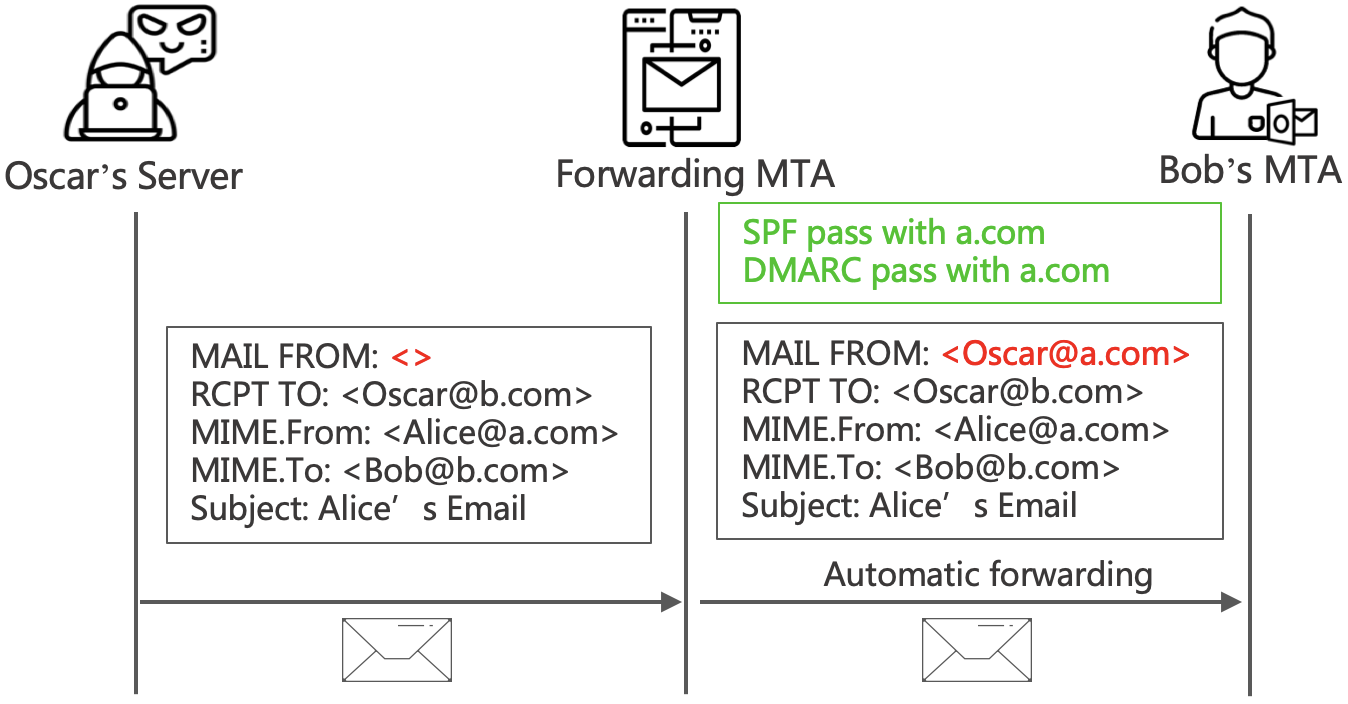}
\caption{Exploiting forwarding services to bypass SPF and DMARC.}
\label{fig:forwarding1}
\end{figure}

% \myparagraph{} 
\noindent \textbf{Unauthorized Forwarding Attack (A$_{9}$)\label{A9}.}
If the attacker can freely configure forwarded emails to any accounts without any authentication verification, the email service has unauthorized forwarding issues. 
First, the attacker should have a legitimate email account on the email forwarding service. 
Because these emails are sent from a well-known email forwarding MTA, the receiver's MTA generally accepts such emails. 
We can also exploit forwarding services to bypass SPF and DMARC protocols when the target domain name is the same as the forwarding domain name.
This attack is depicted in Figure~\ref{fig:forwarding1}. 
Based on this attack, attackers can abuse the credibility of well-known MTAs to craft an realistic spoofing email.

Among our experimental targets, 12 email services have such vulnerabilities.
7 email services do not provide the email forwarding feature.
The other email services have realized the risks and performed corresponding forwarding verification to fix it.

% \myparagraph{}
\noindent \textbf{The DKIM Signature Fraud Attack (A$_{10}$)\label{A10}.}
The forwarding service may give the forwarded email a higher security endorsement. 
But this feature can be abused by the attacker to send spoofing emails.
The forwarder should not add a DKIM signature of its domain name if the forwarded email does not have a DKIM signature or fails the DKIM validation before. 
%Otherwise, the attacker can defraud the legitimate DKIM signature in this way. 
Otherwise, the attacker can defraud the forwarding services of legitimate DKIM signature. 
However, both RFC 6376\cite{mailsignatures} and RFC 6377\cite{kucherawy2011domainkeys} suggest that forwarders should add their signatures to the forwarded emails. 
It has further led to more email services have such problems.

\begin{figure}[t]
\centering
\subfigure[The spoofing email defraud a DKIM signature signed by a.com.]{
\includegraphics[width=0.45\textwidth]{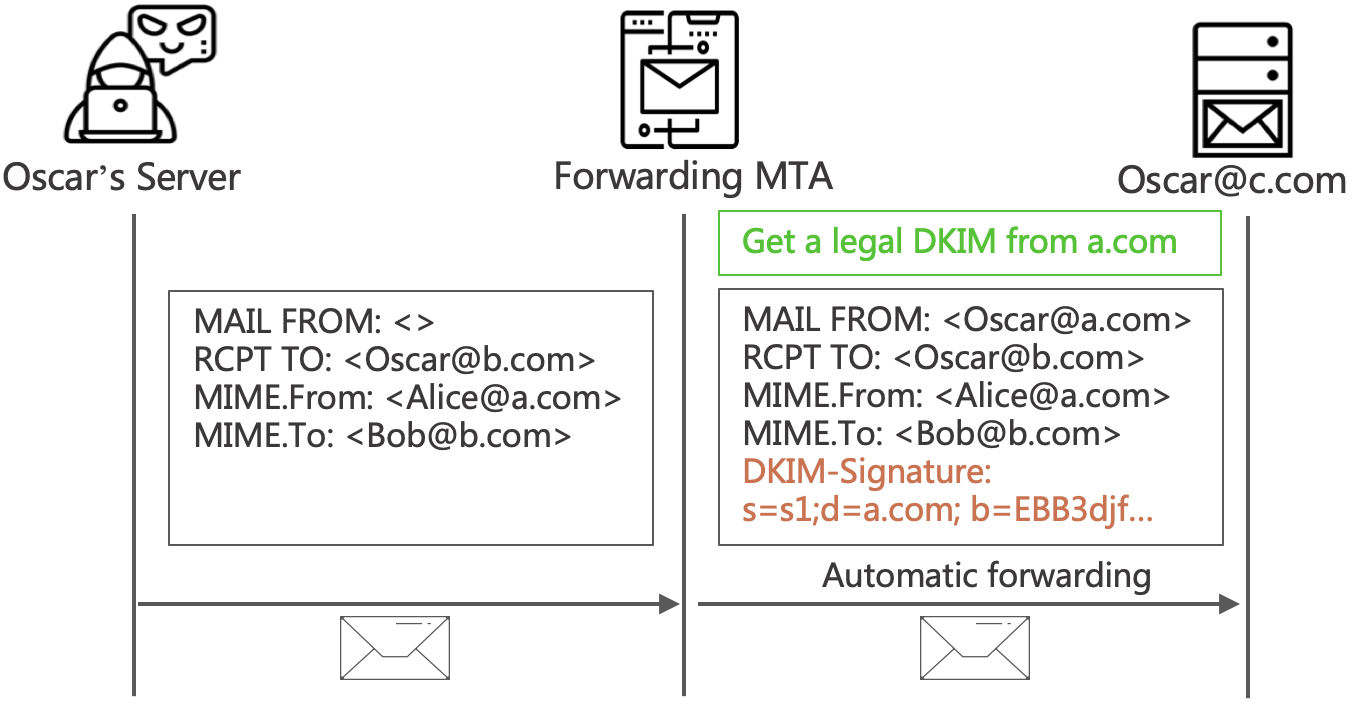}
\label{fig:forwarding2-1}}
\subfigure[Spoofing with the legal DKIM signature.]{
\includegraphics[width=0.3\textwidth]{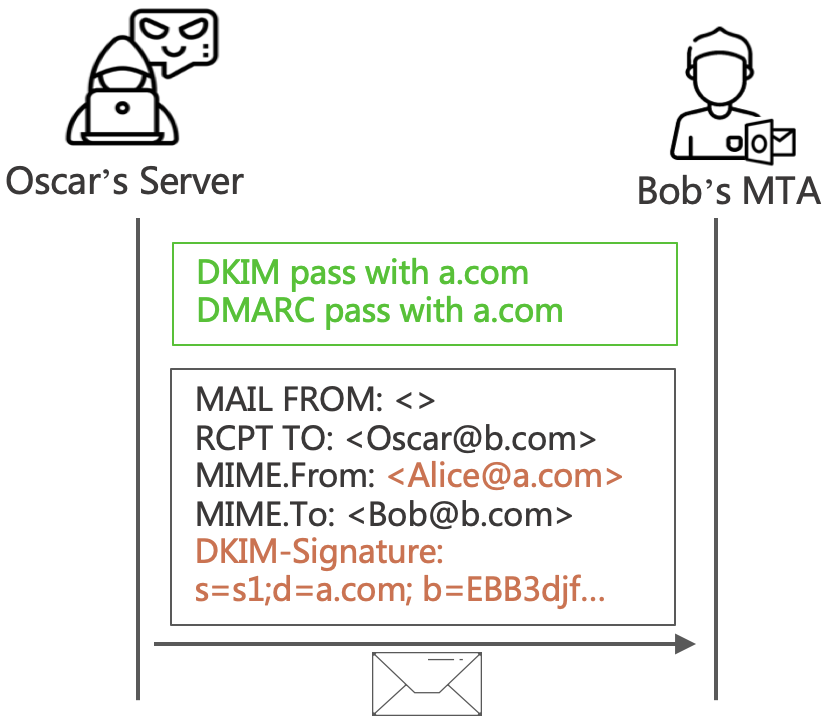}
\label{fig:forwarding2-2}}
\caption{Exploiting forwarding services to bypass DKIM and DMARC.}
\label{fig:forwarding2}
\end{figure}

Figure~\ref{fig:forwarding2} illustrates the complete process of the attack. 
The email forwarding service (a.com) signs and adds DKIM signatures to all forwarded emails without strict verification. 
First, the attacker can register an account (Oscar@a.com) under the email forwarding service. 
Second, he can configure all receiving emails forward to another attacker's email address (Oscar@c.com).
The attacker can then send a spoofing email with \texttt{From}: Alice@a.com, \texttt{To}: Bob@b.com to Oscar@a.com through the direct MTA attack model. 
The forwarding service (a.com) adds a legal DKIM signature to this spoofing email. 
As a result, the attacker gets a spoofing email with a legal DKIM signature signed by a.com.
In our experiments, Alibaba Cloud, Office 365, and Yahoo Email are all vulnerable to such attacks.

% \myparagraph{}
\noindent \textbf{ARC Problems (A$_{11}$)\label{A11}.}
ARC\cite{blank2019authenticated} is a newly proposed protocol that provides a chain of trust to link the verification results of SPF, DKIM, and DMARC in the email forwarding process. 
Only three email services (i.e., Gmail, Office 365, and Zoho) deploy the ARC protocol in our experiments. 
However, our research found that both Office 365 and Zoho have security issues with the ARC protocol implementation. 
Besides, except for the A$_{10}$ attack, ARC cannot defend against most of the attacks discussed above.

For Zoho email services, it shows alerts for users if the email fails the sender inconsistency checks. 
However, there is an error in Zoho's ARC implementation. 
When a spoofing email is automatically forwarded to the Zoho mailbox via Gmail, the ARC-Authentication-Results (AAR) header added by Zoho shows a wrong "pass" DMARC verification result. 
Even worse, this incorrect ARC implementation can also bypass the sender inconsistency checks. 
Zoho does not display alerts to users for this spoofing email.
Office 365 also has errors in the implementation of ARC. 
It passes the wrong verification results of SPF, DKIM, and DMARC in the AAR header. 
This would break the ARC trust chain, which introduces more security risks.

\subsection{Attacks in Email UI Rendering}

The last and most crucial part of the email system is to ensure that emails are rendered correctly. 
Once the attacker can break the defensive measures in this stage, ordinary users are easily deceived by such spoofing emails unconsciously.

The displayed address is the sender address shown on the MUA, but the real address is 
the sender identity (\texttt{From}) used in SMTP communication. 
If an attacker can make the displayed address inconsistent with the real address, the attack is considered successful.
Besides, as shown in Figure ~\ref{fig:sender-check}, some MUAs add a security indicator to those emails which fail the sender inconsistency checks. 
If an attacker can bypass the sender inconsistency checks, it is also regarded as an effective attack technique.

There are various attacks in the email UI rendering stage. 
Some are similar to the  A$_6$, A$_7$ attacks discussed previously. 
The difference is that a UI level attack's goal is to bypass the sender inconsistency checks and spoof the email address shown for users, rather than bypass the three email security protocols' verification. 
Thus, we usually construct ambiguous \texttt{From} headers rather than \texttt{Mail From} headers.
In this section, we only discuss the attack techniques not previously mentioned.

% \myparagraph{}
\noindent \textbf{IDN Homograph Attack (A$_{12}$)\label{A12}.}
The homograph attack\cite{gabrilovich2002homograph} is a known web security issue, but its security risks to the email system have not been systematically discussed. 
As popular email providers gradually support the emails from internationalized domain names (IDN), this attack is likely to have a wider security impact. 

\begin{figure}[t]
\centering
\includegraphics[width=8cm]{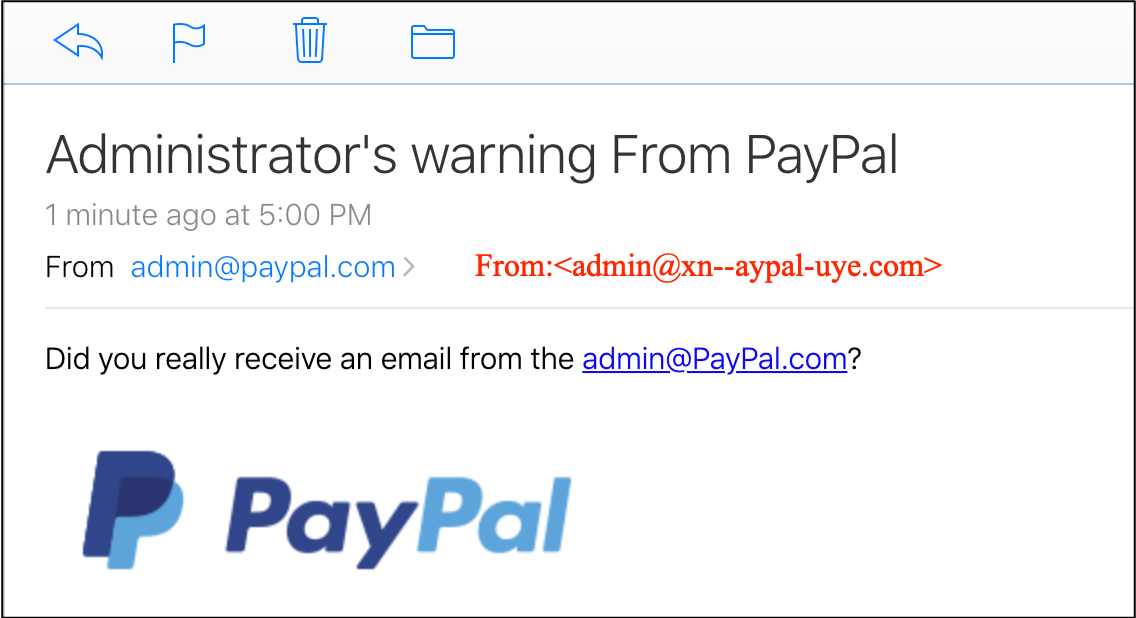}
\caption{A example of IDN homograph attack to impersonate $admin@paypal.com$ on iCloud.com web interface.}
\label{fig:IDN_attack}
\end{figure}

Punycode is a way of converting words that cannot be displayed in ASCII into Unicode encoding. 
Notably, Unicode characters can have a similar appearance on the screen while the original addresses are different.
Figure \ref{fig:IDN_attack} shows a spoofing email that seems to come from the address (admin@paypal.com), but is actually from the address (admin@xn--aypal-uye.com).

Modern browsers have implemented some defensive measures against the IDN homograph attack.
For example, the IDN should not be rendered if the domain label contains characters from multiple languages. 
Unfortunately, we found few similar defensive measures in email systems.

The experimental results show that 10 email services (e.g., Gmail, iCloud, Mail.ru) support IDN email is displayed. 
Currently, only Coremail fixes this vulnerability. 
With our assistance, Coremail adds white spaces before and after the Unicode characters in the address bar. 
In this way, users can easily distinguish between ASCII characters and Unicode characters to prevent such attacks.

% \myparagraph{}
\noindent \textbf{Missing UI Rendering Attack  (A$_{13}$)\label{A13}.}
We also find that many characters can affect the rendering of the MUA. 
Some characters may be discarded during the rendering process. 
Additionally, some characters may also cause the email address to be truncated (similar to the attack A$_6$).  
These characters include invisible characters (\texttt{U+0000–U+001F,U+FF00-U+FFFF}) and semantic characters (\texttt{@,:,;,"}).
For example, the MUA renders the address \texttt{admin@gm@ail.com} as \texttt{admin@gmail.com}.

There are still 12 email services (e.g., zoho.com, 163.com, sohu.com) vulnerable to such attacks. 
Other services refuse to receive or just throw such emails into the spam box.

% \myparagraph{}
\noindent \textbf{Right-to-left Override Attack (A$_{14}$)\label{A14}.}
Several characters are designed to control the display order of the string.
One of these is the "RIGHT-TO-LEFT OVERRIDE" character, U+202E which tells computers to display the text in a right-to-left order. It is mainly used for writing and reading Arabic or Hebrew text.  
Although this attack technique\cite{bidirection} has been discussed elsewhere, its security risk to email spoofing has not yet been fully explored.
An attacker can construct a string as \texttt{$\backslash$u202emoc.a@$\backslash$u202dalice}, which is displayed as \texttt{Alice@a.com}.
Because spoofing emails with RTL characters may be directly thrown into the spam box, we generally encode the payload (with utf-8 mode) to attack.

11 email services (e.g., Outlook, Yahoo, Yandex) are still vulnerable to this attack. 
10 services (e.g., cock.li,daum.net, onet.pl) cannot correctly render this type of email address.
Other email services directly reject such mails.

%% file: 5_combined_attacks.tex
\section{Combined Attacks\label{combined_attacks}}

\replaced[id=skw]{According to four authentication stages in email delivery process, we divide our attacks into four categories.}{
As mentioned above, we divide our attacks into four categories according to the four authentication parts in the email delivery process.}
However, these attacks have certain limitations.
\replaced[id=skw]{
First, some attacks (e.g., A$_2$, A$_3$) can have a spoofing effect on the recipent. However, they can not bypass all email spoofing protections.
}
{First, some attacks (e.g., A$_2$, A$_3$, A$_7$) can not bypass all email spoofing protections, while they can already have a spoofing effect on the recipient.}
For example, a spoofing email via Empty Mail From Attack (A$_3$) bypasses the SPF verification but fails in the DMARC verification. 
\replaced[id=skw]{
In addition, most email vendors have fixed the individually conducted attacks which can bypass all the three email security protocols in our experiment.}{
In addition, most email vendors have fixed those attacks (e.g., A$_1$) which can be conducted individually to bypass all three email security protocols in our experiment.}
Thus, combining multiple attacks of different stages is more feasible \replaced[id=skw]{in practice}{in the real world}.
With a "cocktail" joint attack combining different attack techniques, we can easily construct a spoofing email that can completely pass the verification of three email security protocols and user-interface protections. 
%Besides, 
Finally, there is no difference shown on the receiver's MUA between this spoofing email and a legitimate one. 

There are numerous feasible combined attacks by combining 3 types of attack models and 14 attack techniques in the 4 authentication stages.
This work selects two of the most representative examples to illustrate the effects of combined spoofing attacks.
Table \ref{tab:case_study} lists key information of the two examples.

\begin{table*}[]
\caption{Details of two combined attack examples.}
\label{tab:case_study}
\centering
\begin{threeparttable}
\setlength{\tabcolsep}{1mm}{
\begin{tabular}{ccccccc}
\toprule
\textbf{Attack} & \textbf{From} & \textbf{To}  & \textbf{Attack Model}      & \textbf{Combination of attacks} \\ %& SPF  & DKIM & DMARC \\
\midrule %表中直线
\textbf{Case 1} & admin@paypal.com & victim@icloud.com & Shared MTA Attack & A$_2$ + A$_4$        \\        %& pass & pass & pass \\ 
\textbf{Case 2} & admin@aliyun.com & victim@gmail.com  & Direct \& Forward MTA Attack  & A$_2$+A$_3$+A$_{10}$ \\     %& pass & pass & pass  \\
\bottomrule
\end{tabular}
}
\end{threeparttable}
\end{table*}

\begin{figure}[t]
\centering
\includegraphics[width=3.4in]{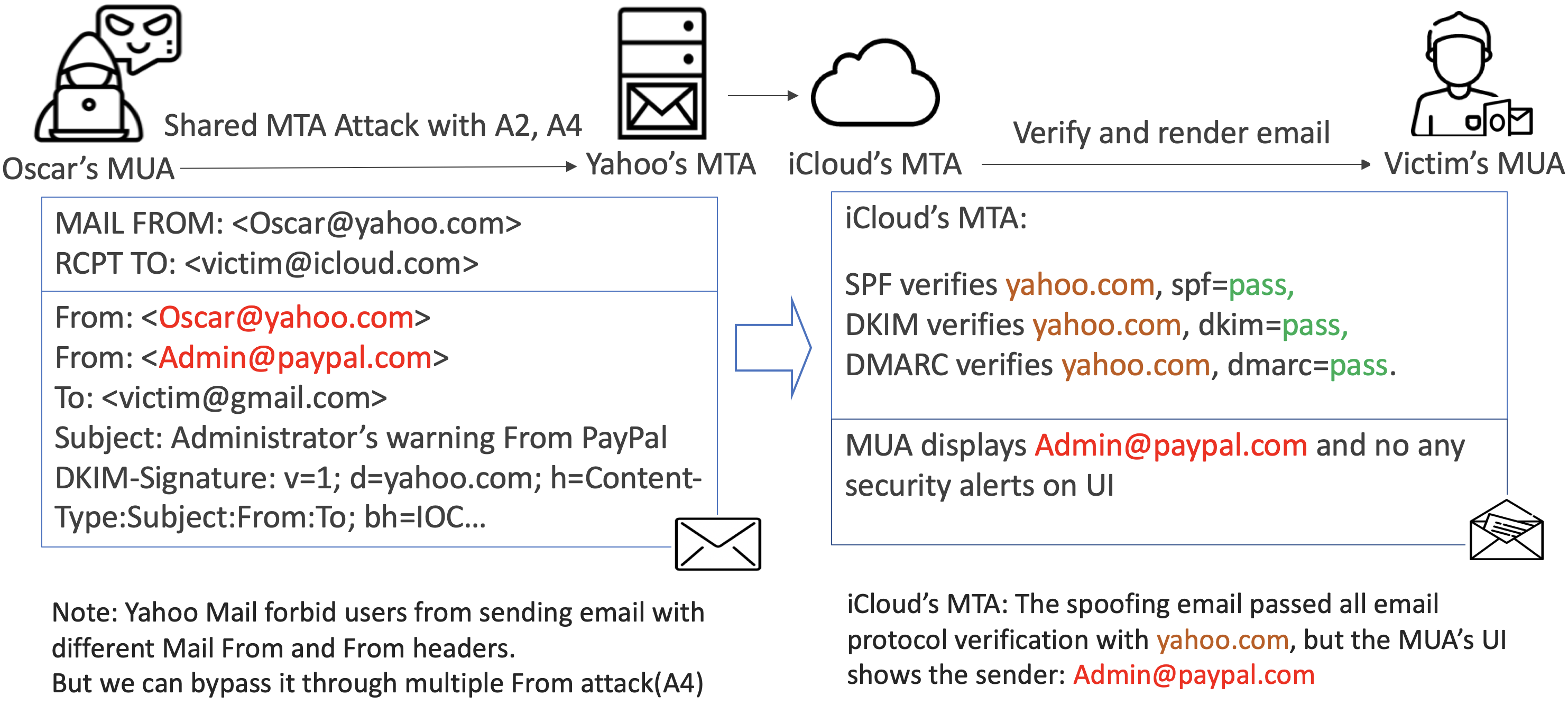}
\caption{Combining A$_2$ and A$_4$ attacks to impersonate $admin@paypal.com$ on iCloud.}
\label{fig:yahoo_icloud}
\vspace{-0.02cm}
\end{figure}
    
\noindent \textbf{Combined Attacks under the Same Attack Model.} 
We identified a total of 14 email spoofing attack techniques, of which 14 attack techniques 
can be combined under the same attack model to achieve better attack effects. 
In addition, although some vendors might fix a vulnerability through one security check, the attacker can accurately combine other attack techniques to bypass the corresponding security check.

Figure~\ref{fig:yahoo_icloud} shows a representative example under the shared MTA attack model. 
Yahoo email performs a simple sender check policy to defend against the A$_2$ attack.  
It prohibits user from sending emails with different \texttt{Mail From} and \texttt{From} headers.  
However, the attacker can still bypass this sender check policy through the A$_4$ attack. 
\replaced[id=skw]{To be specific}{In details}, we can send a spoofing email with a first \texttt{From} header  (\texttt{Oscar@yahoo.com}), which is same as the \texttt{Mail From} header. 
Then, we add a second \texttt{From} header ( \texttt{Admin@paypal.com}). 
Interestingly, iCloud does not reject such a spoofing email with multiple \texttt{From} headers. 
Even worse, iCloud uses the first \texttt{From} header to perform the DMARC verification and gets a "pass" result with yahoo.com, while the second \texttt{From} (Admin@paypal.com) header is displayed on the webmail's UI for users. 
Therefore, this combined attack can eventually bypass all three email security protocols and spoof the MUA.

\noindent \textbf{Combined Attacks under Different Attack Models.} 
The attacker can also conduct a more effective attack by combining different attack models. 
The email system is a complex ecosystem with a multi-party trust chain, which relies on security measures implemented and deployed by multiple parties.
\replaced[id=skw]{Under}{When faced with} different attack models, multiple parties may have various vulnerabilities.  
For example, it is difficult to attack through the shared MTA attack model if a email service's sending MTA performs strict checks in sending authentication. 
However, once it fails to provide a correct and complete security defensive solution in other stages, the attacker can still bypass and send spoofing emails through the other two attack models. 
Hence, we have more combination attacks in the real world by combining multiple attack models.

\begin{figure}[t]
\centering
\subfigure[The first stage of the attack obtained an Alibaba Cloud legal DKIM signature.]{
    \begin{minipage}[t]{1\linewidth}
        \centering
        \includegraphics[width=3.4in]{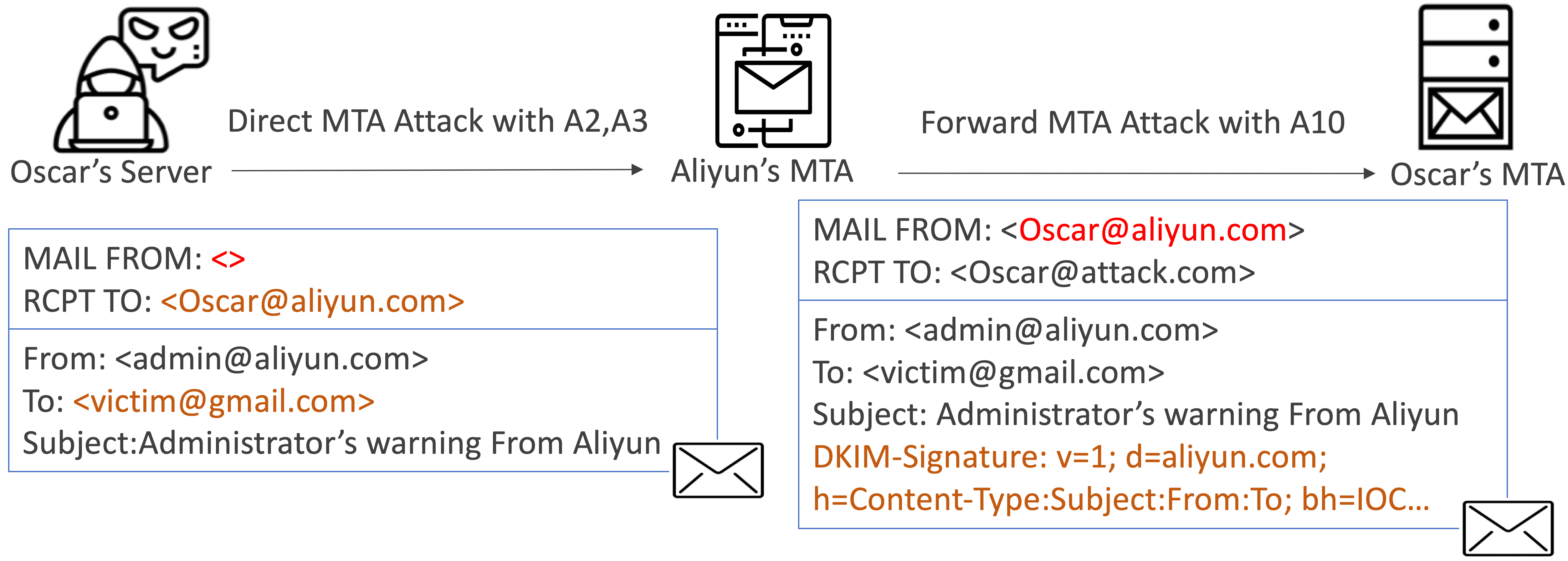}\\
        \vspace{0.02cm}
		\label{fig:combined_attack1_1}
    \end{minipage}
}%
\quad
\subfigure[The second stage of the attack passed Gmail's three mail protocol security verifications.]{
    \begin{minipage}[t]{1\linewidth}
        \centering
        \includegraphics[width=3.4in]{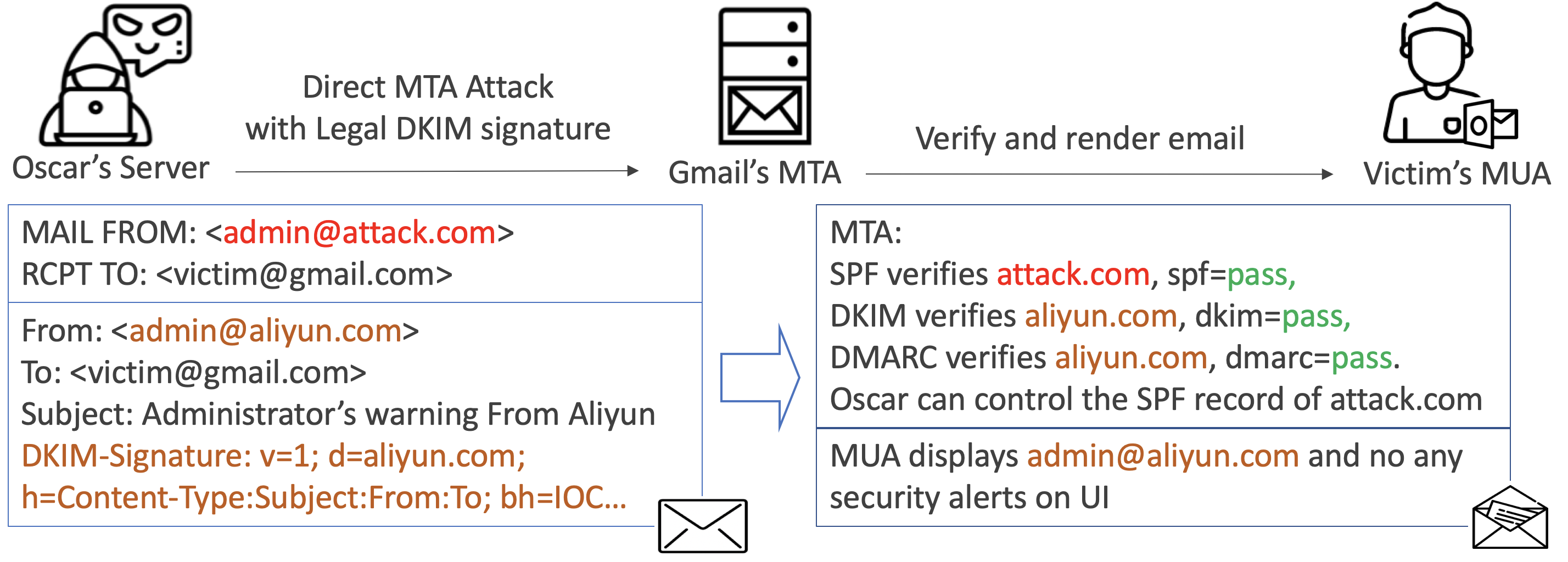}\\
        \vspace{0.02cm}
        \label{fig:combined_attack1_2}
    \end{minipage}%
}%

\caption{A combination attack with A$_2$,A$_3$ and A$_{10}$ from admin@aliyun.com to victim@gmail.com.}
\vspace{-0.2cm}
\label{fig:combined_attack1}
\end{figure}

Figure~\ref{fig:example_attack} shows a successful spoofing attack by combining the direct and forward MTA attack models. 
\replaced[id=skw]{For instance}{In details}, Oscar employs the attack techniques (A$_2$,A$_3$) to send a spoofing email with empty \texttt{Mail From} and crafted \texttt{From} headers. 
Besides, Oscar has a legitimate account (Oscar@aliyun.com), which is different from the victim's account. 
Thus, Oscar can configure this account to automatically forward the received emails to one of his accounts (Oscar@attack.com).
Alibaba Cloud service adds a DKIM signature to all forwarded emails without a necessary verification check (A$_{10}$). 
It grants Oscar's spoofing email a legitimate DKIM signature. 
Then, Oscar can send this spoofing email with \texttt{Mail From:<admin@attack.com>} header through the direct MTA attack model, which is illustrated in  Figure~\ref{fig:combined_attack1_2}.

For this spoofing email, the SPF protocol verifies the \texttt{attack.com} domain, while the DKIM and DMARC protocols verify the \texttt{aliyun.com} domain. 
Therefore, this email can pass all the three email security protocols, and enter the inbox of Gmail. 
In addition, no email service shows alerts for users about the email 
\replaced[id=skw]{with different verified domains of the three protocols.}{whose three protocols' verified domains are different. }
It further makes this type of attack more deceptive to ordinary users.

%% file: 6_discussion_mitigation.tex
\section{Root Causes and Mitigation \label{discussion}}
\label{sec:discussion_mitigating measures}

\subsection{Root Causes}

As aforementioned, the security of email systems  relies on several protection policies that are separately enforced by multiple parties.
Thus, the inconsistencies in these multiple parties could create more vulnerabilities and lead to severe spoofing attacks. 
We identify the root causes of the attacks as follows.

\noindent \textbf{Weak Links among Multi-protocols.}
The protocol verification process is one of the weak links in the authentication chain, due to the ambiguity of email specifications, the lack of best practice and the complexity of the MIME standard.
In the SMTP communication process, multiple fields of protocols contain sender's identity information (i.e., \texttt{Auth username}, \texttt{MAIL From}, \texttt{From}, \texttt{Sender}). The inconsistency of these fields provides the basis for email spoofing attacks. 

SPF, DKIM, and DMARC are proposed and standardized to prevent email spoofing attacks from different aspects. 
However, an email system can prevent email spoofing attacks only when all protocols are well enforced.
In this chain-based authentication structure, a failure of any link can render the authentication chain invalid.

\noindent \textbf{Weak Links among Multi-roles.}
In the email system, authenticating the sender's identity is a complicated process.
It involves four important roles: senders, receivers, forwarders, and UI renderers.
Standard security models work on the assumption that each role properly develops and implements related security verification mechanisms to provide the overall security.
However, many email services do not implement the correct security strategy in all four roles.

Many email services (e.g., iCloud, Outlook, Yeah.com) do not notice the security risks caused by unauthorized forwarding attacks (A$_{9}$) in the email forwarding stage. 
In addition, the specifications do not state any clear responsibilities of four roles (i.e., senders, receivers, forwarders, and UI renderers) in email security verification.

\deleted{
Some services (e.g., Gmail, Yandex.com) forbid sending emails with ambiguous headers but receive them with tolerance. 
Conversely, some (e.g., Zoho, Yahoo) tend to allow the sending of emails with an ambiguous header, but conduct very strict checks in the email receiving verification stage. 
The differences of security policies allow attackers to send spoofing emails from a service with a tolerant sending policy to a service with a loose receiving strategy.
}

\noindent \textbf{Weak Links among Multi-services.}
Different email services usually have different configurations and implementations. 
\added{
Some services (e.g., Gmail, Yandex.com) forbid sending emails with ambiguous headers but receive them with tolerance. 
Conversely, some (e.g., Zoho, Yahoo) tend to allow the sending of emails with an ambiguous header, but conduct very strict checks in the email receiving verification stage. 
The differences among security policies allow attackers to send spoofing emails from a service with a tolerant sending policy to a service with a loose receiving strategy.
}
\deleted{
Some email services are set up in a complicated environment, such as collaborating with third-party service providers. 
It is hard to ensure that every email can pass all SPF, DKIM, and DMARC protocol verification, especially when these environments are in perpetual flux.
}

Besides, some email providers deviate from RFC specifications while dealing with emails with ambiguous headers.
When MUA handles with multiple \texttt{From} headers, some services (e.g., Outlook,Mail.ru) display the first header, while others (e.g., iCloud, yandex.com) display the last header. 

Moreover, different vendors support Unicode characters to various degrees. 
Some vendors (e.g., 21cn.com, Coremail) have been aware of the new security challenges caused by Unicode characters, but some (e.g., 163.com, yeah.net) have no knowledge.
Particularly, some (e.g., zoho.com, EwoMail) even have not yet supported Unicode characters' rendering.

Finally, only a few email providers show visual UI notification to alert users of spoofing emails and only 12 vendors implement sender inconsistency checks. 
In particular, the sender inconsistency checks in practice are significantly diverse because of the absence of a unified implementation standard.
The lack of an effective and reasonable email security notification mechanism is also one reason why email spoofing has been repeatedly prohibited, but never eliminated.

\subsection{Mitigation}

% So far, this work has explored the scope and root causes of these attacks. 
This subsection discusses the key mitigating measures. 
Since email spoofing is a complex problem involving multiple parties, multi-party collaboration is required to counter the relevant issues.

\noindent \textbf{More Accurate Standard.} 
\replaced[id=skw]{Note that email providers may fail to offer a secure and reliable email service with ambiguous definitions in email protocols.}{
Note that, if the email protocol has a too ambiguous definition, email providers may fail to offer a secure and reliable email service.} 
Thus, providing more accurate email protocol descriptions is necessary to eliminate inconsistencies in the practice of multi-party protocols.
For example, the DKIM standard should specify when a DKIM signature should be added to forwarded emails. 
It is reasonable for forwarders to add DKIM signatures to improve the credibility of emails; however, they should not add DKIM signatures to emails that have never passed DKIM verification.

\noindent \textbf{UI Notification.}
Email UI rendering is a significant part that affects the users' perception of an email's authenticity.
Unfortunately, most of webmails and email clients in our experiments \replaced[id=wch]{only show the \texttt{From} header without any more authentication details.}{do not explicitly display SPF, DKIM, or DMARC authentication results.}
\replaced[id=wch]{Therefore, it is difficult for ordinary users to judge the authenticity of emails.}{Thus, it is difficult for most users to understand the details of email authentication fully.}

Additionally, some visual attacks (e.g., A$_{12}$, A$_{13}$) can not be defended at the protocol level. 
An effective defense method is to provide a user-friendly UI notification and alerts users that their received emails may be spoofing emails.
Hu et al.~\cite{hu2018end} also demonstrate that a good visual security notification has a positive effect on mitigating phishing email threats in the real world.
As shown in Figure~\ref{fig:example_attack}, the spoofing email in Section~\ref{combined_attacks} can be verified by all the three email protocols.
\replaced[id=skw]{Nevertheless, users can not distinguish this spoofing email from normal emails without a UI notification.}{
Without a UI notification, users can not see any difference between this spoofing email and normal emails. }

\begin{figure}[t]
\centering
\includegraphics[width=3.2in]{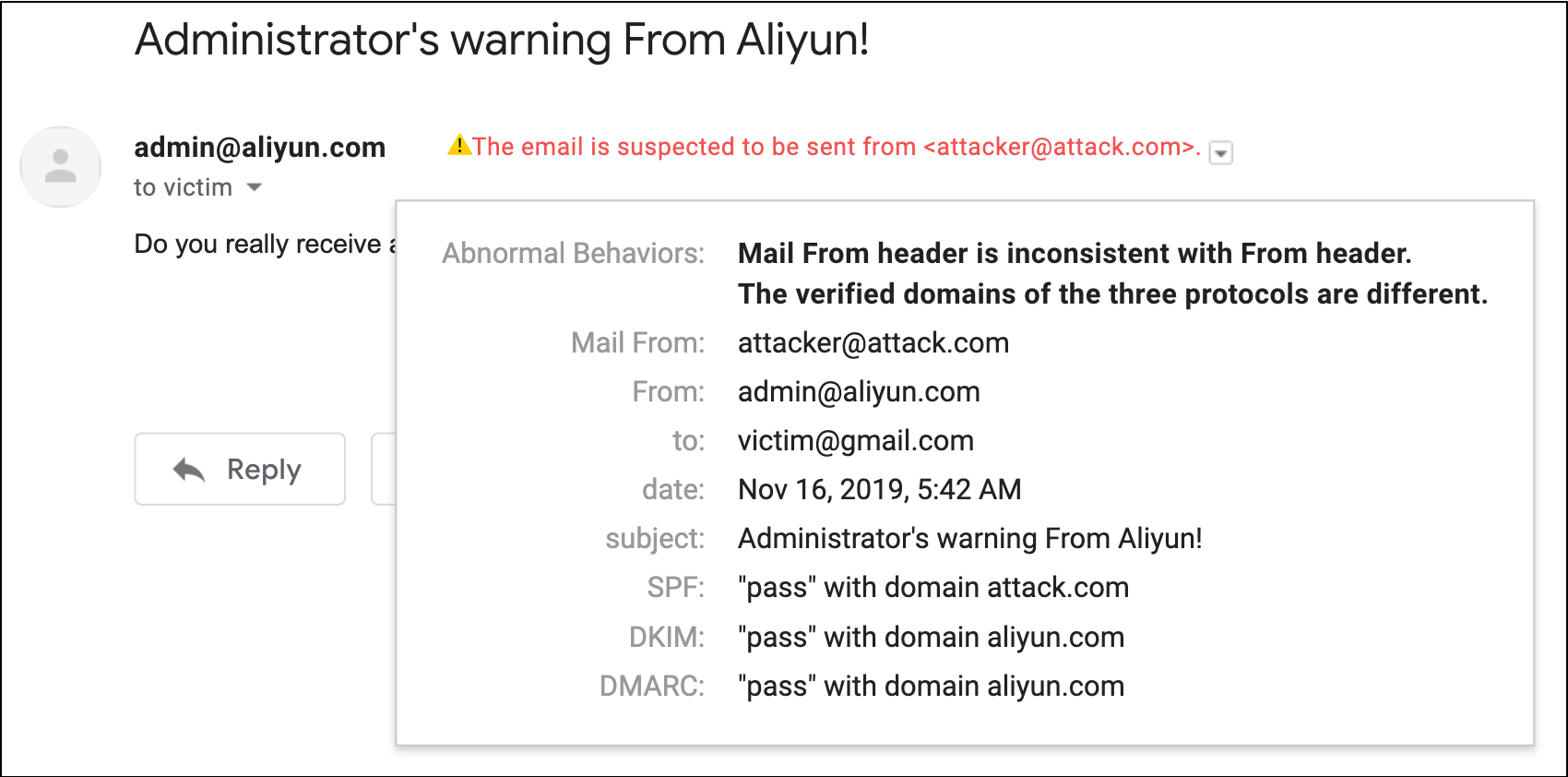}
\caption{An example of UI notification against the combined attack }
\label{fig:ui_demo}
\vspace{-0.02cm}
\end{figure}

As shown in Figure~\ref {fig:ui_demo}, users intuitively can recognize whether the received email contains malicious behaviors, based on the UI notification. 
Coremail, a well-known email service provider in China, has adopted our suggestions and implemented the UI notification on its webmail and email client. In addition, we have released the UI notification scheme in the form of a chrome extension for Gmail called "NoSpoofing"\footnote{NoSpoofing : \url{https://chrome.google.com/webstore/detail/nospoofing/ehidaopjcnapdglbbbjgeoagpophfjnp}}.

\noindent \textbf{Evaluation Tools.} 
We have released our testing tool publicly on GitHub \footnote{Email Spoofing Test Tool: \url{https://github.com/mo-xiaoxi/EmailSpoofingTestTool}} for email administrators to evaluate and increase their security. 
After configuring the target email system information, the tool can interact with the target system and evaluate whether the target system is vulnerable to the attacks. 
%For the vulnerable attacks, administrators can configure corresponding filtering rules to defend against these attacks.

%% file: 7_disclosure_and_response.tex
\section{Disclosure and Response}
\label{sec:disclosure}

Vulnerabilities found in this work have already been reported to all 30 relevant email vendors in detail. 
We have been contacting these entities to help them mitigate the detected threats. 
Our contact results are summarized as follows.
%We summarized the results of our contacts as below.
%We attempted to contact all 30 email vendors in two main ways. We provided detailed reports to email vendors' security contacts. We also reported the problem to CNCERT/CC and the CERT coordination center (CERT/CC) through the HackerOne platform. Below we summarize the discussions.

\noindent \textbf{Alibaba Cloud:} 
They are interested in the attacks and have an in-depth discussion with us about the specifications. 
They mention that RFC 6376 suggests adding a DKIM signature in the email forwarding stage to increase emails' credibility.
They have now recognized the risk of adding DKIM signatures without verification and promise to evaluate and fix such issues. 
They also suggest we contact the authors of related RFCs to reach an agreed fix proposal.

\noindent  \textbf{Gmail:} They acknowledge our report and will fix related issues in subsequent updates. 
They contact us for discussing the essential reasons behind these security issues.

\noindent \textbf{iCloud:} They discuss with us about the details of the attacks and their potential consequences.
In particular, Apple iCloud Email has already fixed related security issues with our cooperation.

\noindent \textbf{Sina:} They evaluate the issue as a high-risk vulnerability and internally assess the corresponding protective measures. As a bonus, they provide us a reward of $ \approx $ \$90.

\noindent \textbf{Yandex:} They accept our report and confirm the vulnerability. At the same time, they provide a bonus of \$200 for appreciation.

\noindent \textbf{Yahoo:} They confirm the vulnerability. But they claim that it is not an immediate risk.

\noindent \textbf{Coremail:} They acknowledge our report and particularly thank us for reporting the issue of UI attacks. To counter those security issues, they adopt our suggestions and and start to implement the UI notification to protect users against email spoofing attacks. 

\noindent \textbf{QQ Mail and 163.com:} They appreciate our work and inform us that they would fix those security issues by anti-spam strategies.

\noindent \textbf{Outlook and Mail.ru:} They claim that they are strictly operating their email service in accordance with RFC standards. They categorize these problems as phishing emails and promise to pay more attention to the impact of such attacks.

\noindent \textbf{Others:} We have contacted other relevant email vendors and look forward to receiving their feedback. 

%\textbf{RFC:}

%% file: 8_related_work.tex
\section{Related Work}
\label{sec:related}

% Recently, various studys have highlighted the threat of email phishing and also explored potential defenses.

%\subsection{Email Phishing}
Prior works have revealed certain threats of phishing email attacks\cite{conway2017qualitative,drake2004anatomy}, including the impacts of spear phishing attacks on email user's behavior\cite{lin2019susceptibility}. 
Our work focuses on more novel forms of spoofing attacks and their influence on the whole authentication process.
Poddebniak et al.~\cite{poddebniak2018efail} discuss how practical spoofing attacks break various protections of OpenPGP and S/MIME email signature verification. 
They also discuss two new protocols that are proposed to enhance spoofing detection, such as BIMI (Brand Indicators for Message Identiﬁcation)~\cite{bimi2019} and ARC (Authenticated Received Chain)~\cite{andersen2017authenticated}. 
However, BIMI is built on DMARC and has not been fully standardized. 
Thus, the attacks we found are also effective.
ARC protocol is standardized in 2019, yet, only three vendors (i.e., Gmail, Office 365, Zoho) have deployed the protocol in our experimental targets.
Our work finds that, however, both Office 365 and Zoho have flaws with the implementation of ARC, which can still lead to some security issues .

Hu et al.~\cite{hu2018end} analyzed how email vendors detect and handle spoofing emails through an end-to-end email spoofing experiment. We find that the vulnerabilities they mentioned have been mostly fixed in the past two years. Besides, they did not discuss bypassing security protocols detection. 
Our work focuses on new attacks that can bypass security protocols or user-interface protections. We can construct a highly realistic spoofing email that can completely bypass all the email security protocols and user-interface protections.
%Our work is complementary to existing work to depict a complete picture.

%\subsection{Defense}
In addition, prior literature has proposed many techniques to defend traditional phishing attacks. 
SMTP extensions, such as SPF, DKIM, and DMARC, are designed to protect the authenticity of emails. 
Foster et al. \cite{foster2015security} measured the implementation and deployment of these protocols and pointed out that, unfortunately, despite years of development, the acceptance rate of these security protocols are still not very high. 
This low acceptance rate seriously jeopardizes the security of the email ecosystem\cite{hu2018towards}. 

Besides, there are many works discussing phishing detection methods based on features extracted from email content and headers\cite{fette2007learning,cidon2019high,krause2019recognizing}, lots of which rely on machine learning technology. 
Furthermore, Ho et al.~\cite{ho2017detecting} point out the positive effects of a good security metric against phishing attacks.
Other works\cite{hu2018revisiting,petelka2019put} indicates that the current email services does not have a UI Notification as HTTPS\cite{luo2017hindsight}. 
The contemporary visual security indicators are not enough to provide full phishing protection\cite{hu2018end,krol2012don}. 
For email spoofing attacks, our research provides a UI notification scheme and evaluation tools for email systems' administrators.
It could effectively boost the development of protective measures against email spoofing in the future.

%% file: 9_conclusion.tex
\section{Conclusion}
\label{sec:conclusion}

% Email is a complex ecosystem that relies on a multi-party trust chain structure, whose security mechanisms are implemented by different email services. 
This paper explored the vulnerabilities of the chain-based authentication structure in the email ecosystem. 
Specifically, a failure in any part can break the whole chain under this chain-based structure.
Namely, the authenticity of an email depends on the weakest link in the email authentication chain.

We presented a series of new attacks that can bypass SPF, DKIM, DMARC and user-interface protections through a systematic analysis of the email delivery process. 
In addition, we conducted a large-scale analysis of 30 popular email services and 23 email clients. 
Experiment results show that all of them are vulnerable to the new attacks, including famous email services, such as Gmail and Outlook. 
We underscore the unfortunate fact that many email services have not implemented adequate protective measures. 
Besides, recognizing the limitation of past literature, which focused on spoofing attacks' impacts on a single step of the authentication process, we concentrated on spoofing attacks' influence on the chain-based email authentication process as a whole.

Based on our findings, we analyzed the root causes of these attacks and reported the issues to corresponding email service providers. 
We also proposed key mitigating measures for email protocol designers and email providers to defend against email spoofing attacks. 
Our work is devoted to helping the email industry more efficiently protect users against email spoofing attacks and improve the email ecosystem's overall security.

%% file: 10_acknowledgments.tex
\section*{Acknowlegments}
We sincerely thank our shepherd Zakir Durumeric and all the anonymous reviewers for their valuable reviews and comments to improve this paper.
We also thank Mingming Zhang, Kangdi Cheng, Zhuo Li, Ennan Zheng, and Jianjun Chen for peer-reviewing and assisting in editing this paper.

This work is supported in part by the National Natural Science Foundation of China (Grant No. U1836213 and U1636204), the BNRist Network and Software Security Research Program (Grant No. BNR2019TD01004).